\documentclass[useAMS,usenatbib]{mn2e}
\usepackage{times}
\usepackage{graphicx}
\usepackage{amsmath}
\usepackage{subfigure}
\usepackage{psfrag}
\title[MCMC analysis of Bianchi ${\rm VII_h}$ models]{Markov chain Monte Carlo analysis of Bianchi ${\rm VII_h}$ models}
\author[M. Bridges et al.]
  {M. Bridges,$^1$\thanks{E-mail: m.bridges@mrao.cam.ac.uk}
   J.D. McEwen, $^1$ A.N. Lasenby,$^1$ M.P. Hobson$^1$\\
  $^1$Astrophysics Group,
      Cavendish Laboratory, Madingley Road,
      Cambridge CB3 0HE, UK\\
}

\date{Accepted ---. Received ---; in original form \today}
\pagerange{\pageref{firstpage}--\pageref{lastpage}}
\pubyear{2005}

\begin{document}
\label{firstpage}
\maketitle

\begin{abstract}
We have extended the analysis of \citet{Jaffea, Jaffec} to a complete Markov
chain Monte Carlo (MCMC) parameter space study of the Bianchi type ${\rm VII_h}$
models including a dark energy density, using Wilkinson Microwave Anisotropy
Probe (WMAP) cosmic microwave background (CMB) data from the 1-year and 3-year
releases. Since we perform the analysis in a Bayesian framework our entire
inference is contained in the multidimensional posterior distribution from which
we can extract marginalised parameter constraints and the comparative Bayesian
evidence.  Treating the left-handed Bianchi CMB anisotropy as a template centred
upon the `cold-spot' in the southern hemisphere, the parameter estimates derived 
for the total energy density, `tightness' and vorticity from 3-year data are
found to be:  $\Omega_{tot} = 0.43\pm 0.04$, $h = 0.32^{+0.02}_{-0.13}$,
$\omega = 9.7^{+1.6}_{-1.5}\times 10^{-10}$ with orientation $\gamma =
{337^{\circ}}^{+17}_{-23}$). This template is preferred by a factor of roughly unity in
log-evidence over a concordance cosmology alone.  A Bianchi type template is
supported by the data only if its position on the sky is heavily restricted. All
other Bianchi ${\rm VII_h}$ templates including all right handed models, are disfavoured.
The low total energy density of the preferred template, implies a geometry that
is incompatible with cosmologies inferred from recent CMB observations.
\citet{Jaffeb} found that extending the Bianchi model to include a term in
$\Omega_{\Lambda}$ creates a degeneracy in the $\Omega_m - \Omega_{\Lambda}$
plane. We explore this region fully by MCMC and find that the degenerate
likelihood contours do not intersect areas of parameter space that 1 or 3 year
WMAP data would prefer at any significance above $2\sigma$. Thus we can confirm
the conclusion that a physical Bianchi ${\rm VII_h}$ model is not responsible for this
signature, which we have treated in our analysis as merely a template.

\end{abstract}


\section{Introduction}
It has recently been proposed that some of the current anomalous results seen in
observations of the CMB sky by WMAP, namely non-Gaussian structure centered on
the so called `cold spot' in the southern hemisphere (\citealt{Vielva};
\citealt{Cruz}), low quadrupole \citep{Efstathiou} and multipole alignments
\citep{Oliveira-Costa}, could be removed by allowing for large scale vorticity
and shear. Components such as these arise in Bianchi type ${\rm VII_h}$ models
\citep{Barrow} due to large scale rotation of the universe distorting the CMB
since the time of last scattering. This assumption has the unfortunate
side-effect of violating universal isotropy and hence the cosmological principle,
so any such result should be examined very carefully.

Such a finding was indeed made by \citet{Jaffea} [hereafter referred to as Jaffe]
where a significant correlation was found between a class of Bianchi ${\rm
VII_h}$ models and WMAP observations. However the resultant best-fit Bianchi
component was derived directly from \citet{Barrow} for a universe with a large
negative curvature with total energy density $\Omega_{tot} = 0.5$ and including
no dark energy component. Such a cosmology cannot be reconciled with cosmologies
inferred from either current CMB or other astronomical observations. In
\citet{Jaffeb} the authors extended their study to include a cosmological
constant and searched for a morphologically identical template to their first
analysis in the increased parameter space. We perform a new study to explore
fully the parameter space of both Bianchi and cosmological parameters in order to
select the best fitting template. We carry out this investigation using Markov
chain Monte Carlo (MCMC) sampling in a fully Bayesian manner. Hence we also find
the comparative Bayesian evidence allowing an assessment of whether the inclusion
of such a template is supported by the data. 

\section{MCMC Analysis and Model Selection Framework} 
Several recent studies have been performed using a Bayesian approach to
cosmological parameter estimation and model selection  (\citealt{Jaffe},
\citealt{Drell}, \citealt{John}, \citealt{Slosar}, \citealt{Saini},
\citealt{Marshall}, \citealt{Niarchou}, \citealt{Basset}, \citealt{Mukherjee},
\citealt{Trotta}, \cite{Beltran}, \citealt{Bridges}, \citealt{Parkinson}).  Here
we are interested in determining the best fitting Bianchi ${\rm VII_h}$ sky from
WMAP data using a selection of all sky maps available; the 1-year WMAP Internal
Linear Combination (WILC) map \citep{Bennett}, the Lagrange Internal Linear
Combination (LILC) map \citep{Eriksen}, the Tegmark, de Oliveira-Costa \&
Hamilton (TOH) map \citep{Tegmark}, the Maximum Entropy Method (MEM) map
\citep{Stolyarov} and the new 3-year WILC map \citep{W3ILC}.  We have
incorporated Bianchi sky simulations into an adapted version of the {\sc cosmoMC}
package \citep{cosmomc} which we show via simulated sky maps is capable of
extracting both the underlying cosmology and the Bianchi signal, while also
determining when the addition of such a component is favoured by the data.

The Bayesian framework allows the examination of a set of data $\bf{D}$ under the assumption of a model $M$ 
defined by a set of parameters $\bf{\Theta}$ with Bayes' Theorem:
\begin{equation}
P(\bf{\Theta|\bf{D}, M}) = \frac{P(\bf{D}|\bf{\Theta}, M) P (\bf{\Theta}|
M)}{P(\bf{D}|M)},
\label{Bayes}
\end{equation} 
where $P(\bf{\Theta}|\bf{D},M)$ is the posterior distribution,
$P(\bf{D}|\bf{\Theta},M)$ the likelihood, $P(\bf{\Theta}|M)$ the prior
and $P(\bf{D}|M)$ the Bayesian evidence. We explore the parameter space by MCMC sampling to maximise the
likelihood utilising a reasonable convergence criteria from the  Gelman \& Rubin $R$ statistic \citep{Gelman-Rubin} so that once the Markov chain has reached its 
stationary distribution, samples from it reliably reflect the actual posterior distribution of the model. 
The final result will be parameter estimates marginalised from the multi-dimensional 
posterior and a value of the evidence for the model being studied. 

\subsection{Likelihood}
A CMB experiment records the temperature anisotropy on the sky $\bf{T}(\bf{\hat{p}})$. We wish to find a cosmological model that can
optimally describe these fluctuations by maximising the probability of observing
the data $\bf{T}(\bf{\hat{p}})$ given a model described 
by a set of parameters $\bf{\Theta}_{\rm{C}}$. Then for a Gaussian model, from Bayes' Theorem we can express this probabilty, the \emph{likelihood}, as:
\begin{equation}
P(\bf{T}(\bf{\hat{p}})|\bf{\Theta}_{\rm{C}}) \propto
\frac{1}{\sqrt{|\mathcal{C}|}}e^{-\frac{1}{2}(\bf{T}^{\rm T}\mathcal{C}^{-1}\bf{T})}
\label{eqn1}
\end{equation}
where $\mathcal{C}$ is the covariance matrix containing the predictions of the
model $\bf{\Theta}_{\rm{C}}$ and $|\mathcal{C}|$ 
its determinant. Expanding the temperature sky in spherical harmonmics
$\bf{T}(\bf{\hat{p}}) = \sum_{lm}a_{lm}Y_{lm}$, in a universe
that is globally isotropic means $\mathcal{C}_{lm,l'm'} = \langle a_{lm} a^*_{l'm'} \rangle = C_l \delta_{ll'} \delta_{mm'}$ where the variance
of the harmonic coefficients $a_{lm}$ is $C_l = \frac{1}{2l+1}\sum_m |a_{lm}|^2$. Thus Eqn. \ref{eqn1} becomes
\footnote{Where we have used the notation that $\hat{C_l}$ and $\hat{a_{lm}}$ refer to observed quantities while $C_l$ and $a_{lm}$ refer to their
theoretical counterparts.}:
\begin{equation}
P(\{\hat{a}_{lm}\}|{\bf{\Theta}_{\rm{C}}}) \propto \prod_{lm} \frac{e^{\left(-|\hat{a}_{lm}|^2/2C_l\right)}}{\sqrt{C_l}}. 
\label{eqn2}
\end{equation}
If one assumes universal isotropy the likelihood is then independent of $m$ which can be summed over to give (for a full sky survey) the form used in
most conventional CMB analyses:
\begin{equation}
\ln P(\hat{C_l}|{\bf{\Theta}_{\rm{C}}}) = (2l+1)\left(\frac{\hat{C_l}}{C_l}+\ln|C_l|\right),
\label{fullskylike}
\end{equation}  
where $\hat{C}_l$ is the observed quantity. 
However Bianchi $\rm{VII_h}$ models are anisotropic so one cannot simply compare observed $\hat{C}_l$'s with theoretical $C_l$'s, instead the
full set of $a_{lm}$'s must be used.

In this analysis we make the assumption that there is both a background cosmological contribution and Bianchi $\rm{VII_h}$ contribution 
to the CMB. We take the cosmological component to be that of a standard $\Lambda$CDM universe, producing a set of $a_{lm}$'s, while the
Bianchi model component (defined by parameters $\bf{\Theta}_{\rm{B}}$) is given by $a_{lm}^{\rm{B}}$. 
The Bianchi component can then be \emph{subtracted} from the observed $\hat{a}_{lm}$'s so that Eqn \ref{eqn2}
becomes: 
\begin{multline}
\hspace{-10pt}P(\{\hat{a}_{lm}\}|\Theta_{\rm{B}}, \Theta_{\rm{C}}) \propto \prod_{l} \frac{1}{\sqrt{C_l}}e^{-\left(\frac{(\hat{a}_{l0}-a_{l0}^B)^2}{C_l}\right)} \cdot \\
\prod_{m=1}^{l} \frac{2}{C_l} e^{-\left(\frac{|\hat{a}_{lm}-a_{lm}^B|^2}{2C_l}\right)}.
\end{multline}	 
For numerical convenience we take the natural logarithm to yield the final likelihood
function used in this analysis, hereafter referred to as $\mathcal{L}$:
\begin{multline}
\hspace{-10pt}\ln \mathcal{L} \propto \sum_{l} - (l+1/2)\ln (C_l)\\
\shoveright{ - \frac{(a_{l0}-a_{l0}^{\rm{B}})^2}{2C_l} -\frac{1}{C_l}\sum_{m=1}^{l}\left(|a_{lm}-a_{lm}^{\rm{B}}|^2\right)}, 		
\end{multline}
which would reduce to Eqn. \ref{fullskylike} if the Bianchi signal were zero.

\subsection{Evidence}
The Bayesian evidence is the average likelihood over the
entire prior parameter space of the model:
\begin{equation}
\int\int{\mathcal{L}(\Theta_{\rm{C}},\Theta_{\rm{B}})P(\Theta_{\rm{C}})P(\Theta_{\rm{B}})}d^N\Theta_{\rm{C}} d^M\Theta_{\rm{B}},
\label{equation:evidence}
\end{equation}
where $N$ and $M$ are the number of cosmological and Bianchi parameters respectively.
Those models having large areas of prior parameter space with high likelihoods will produce high evidence values and \emph{vice versa}.
 This effectively
penalises models with excessively large parameter spaces, thus naturally incorporating Ockam's razor.

The results generated in this paper have employed the new method of nested sampling \citep{Skilling} as implemented in a forthcoming publication (Shaw et al. in
preperation). This method is capable of much higher accuracy than previous methods such as thermodynamic integration (see e.g. \citealt{Beltran}, \citealt{Bridges}).
due to a computationally more effecient mapping of the integral in Eqn. \ref{equation:evidence} to a single dimension by a
suitable re-parameterisation in terms of the prior \emph{mass} $X$. This mass can be divided
into elements $dX = \pi(\mathbf{\Theta})d^N \mathbf{\Theta}$ which can
be combined in any order to give say
\begin{equation}
X(\lambda) = \int_{\mathcal{L\left(\mathbf{\Theta}\right) > \lambda}} \pi(\mathbf{\Theta}) d^N
\mathbf{\Theta},
\end{equation}
the prior mass covering all likelihoods above the iso-likelihood curve
$\mathcal{L} = \lambda$.  We also require the function
$\mathcal{L}(X)$ to be a singular decreasing function (which is
trivially satisfied for most posteriors) so that using sampled points
we can estimate the evidence via the integral:
\begin{equation}
\mathcal{Z}=\int_0^1{\mathcal{L}(X)}dX.
\label{equation:nested}
\end{equation}
Via this method we can obtain evidences with an accuracy 10 times higher than previous methods for the same number of likelihood evaluations.

\section{Bianchi ${\rm VII_h}$ Simulations}
\label{bianchi}

Jaffe implemented solutions of the geodesic equations developed by \citet{Barrow} to compute Bianchi
induced temperature maps in type ${\rm VII_h}$ cases. \citet{Jaffeb} extended  these to include cosmologies
with dark energy allowing an almost flat geometry, in keeping with the so called concordance cosmology. One
of us (ANL, in preparation) independently performed the same extension and found results   agreeing well
with \cite{Jaffeb}.

\begin{figure}
\begin{center}
\includegraphics[width=\linewidth]{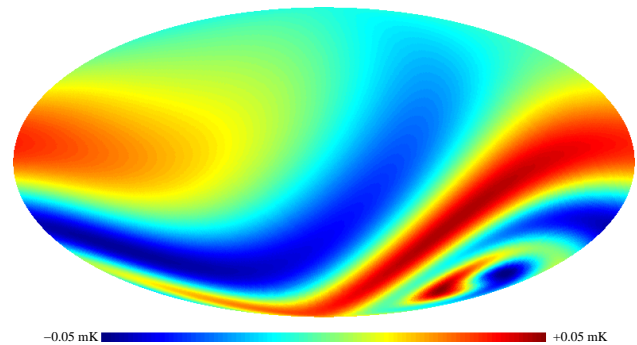}
\caption{Best fit bianchi template as found by \citet{Jaffec} computed from parameters $\Omega_{tot}=0.5$, $\omega=10.38\times10^{-10}$,
$h=0.15125$ and position ($42^{\circ}$, $28^{\circ}$, 
$310^{\circ}$) in Euler angles.}
\label{jaffe best fit}
\end{center}
\end{figure}

\begin{figure}
\begin{center}
\includegraphics[width=\linewidth]{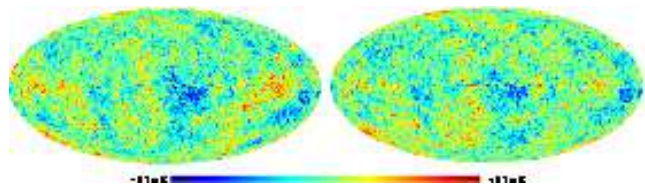}
\caption{Raw LILC map (left) with the Bianchi corrected map (right)}
\label{jaffe_corrected_ilc_comparison}
\end{center}
\end{figure}

The models are parameterised by six independent quantities; the total energy density $\Omega_{tot}$ =
$\Omega_m$ + $\Omega_{\Lambda}$, the current vorticity $\omega$, $h$ which determines the `tightness' of
the spiral and the Euler angles $\alpha$ and $\beta$ determining the position of the centre of the spiral
on the sky with $\gamma$ allowing rotation of the pattern about this point \footnote{We adopt the active
$zyz$ Euler convention corresponding to the rotation of a physical body in a {\it{fixed}} coordinate system
about the $z$, $y$ and $z$ axes by $\gamma$, $\beta$ and $\alpha$ respectively.}. In addition one must
specify the direction of rotation or `handedness' of the spiral.  The ratio of the current shear to the
Hubble parameter $(\sigma/H)_0$ (differential expansion) is determined by the combination of these
parameters: 
\begin{equation*}
\left(\frac{\sigma}{H}\right)_0 = \frac{6h\Omega_{tot}}{\sqrt{2}(1+h)^{1/2}(1+9h)^1/2(1-\Omega_{tot})}\left(\frac{\omega}{H}\right)_0.
\end{equation*}  
Jaffe found a best fit bianchi template $\bf{T}$ by a $\chi^2$ analysis of the observed CMB sky 
$\bf{D}$ for a selection of WMAP data variants by minimising:
\begin{equation}
\chi^2 = (\bf{D}-\bf{T})^T \bf{M}^{-1} (\bf{D} - \bf{T})
\end{equation}
under the assumption of an embedded cosmology $\bf{M}$ given by the best fit 1-year WMAP angular power spectrum \citep{Spergel}. There
is no contribution from noise in this expression as the dominant uncertainty on large scales is from cosmic variance.
Their result is shown in Fig. \ref{jaffe best fit} \footnote{See http://www.mrao.cam.ac.uk/~jdm57 for an animation of this template on the
sphere.} computed in our formalism with parameters $\Omega_{tot}=0.5$, $\omega=10.38 \times 10^{-10}$, $h=0.15125$ and position ($42^{\circ}$, $28^{\circ}$, 
$310^{\circ}$) in Euler angles. The primary feature is a central core positioned on the so called `cold spot', assumed to
be the source of a large area of non-gaussianity. The corrected map (see Fig. \ref{jaffe_corrected_ilc_comparison}), produced after removal of
 this Bianchi `contamination', provides even by cursory visual inspection a more statistically homogeneous
 sky.   

In Section \ref{Simulated Data} we will perform our analysis on simulated data in order to test our parameter recovery and model
selection algorithm with a Bianchi component of varying amplitude superimposed on a background concordance cosmology. In Section
\ref{Real Data} we perform this analysis on real data to  effectively extend Jaffe's study to fit simultaneously for a selection of
right and left handed Bianchi models and a background cosmological component. This method has two main advantages over Jaffe's original
template analysis: i) confirmation of the robustness of the detection within a varying cosmology --thus accounting for possible
degenerate parameters and  ii) estimation of the ratio of Bayesian evidences with and without a Bianchi component --providing an
effective model selection to favour or disfavour the inclusion of a such an addition to the standard model.

\section{Simulated Data}
\label{Simulated Data}
A set of simulated CMB sky maps were produced with the {\sc HEALPix} \footnote{http://healpix.jpl.nasa.gov} software
routines. These contained a cosmic variance limited concordance cosmological component and a Bianchi $\rm{VII_h}$ component with fixed
handedness, shear, matter and dark energy density and geometry but varying the vorticity $\omega$ to study the sensitivity
of the method to the total amplitude  of the Bianchi signal (of which $\omega$ is a good tracer). Evidence estimates
 were produced for two parameterisations: Bianchi + cosmology (model A); and cosmology only (model B). 

\begin{figure}
    	\subfigure[Bianchi]{
          \includegraphics[width=.5\columnwidth]{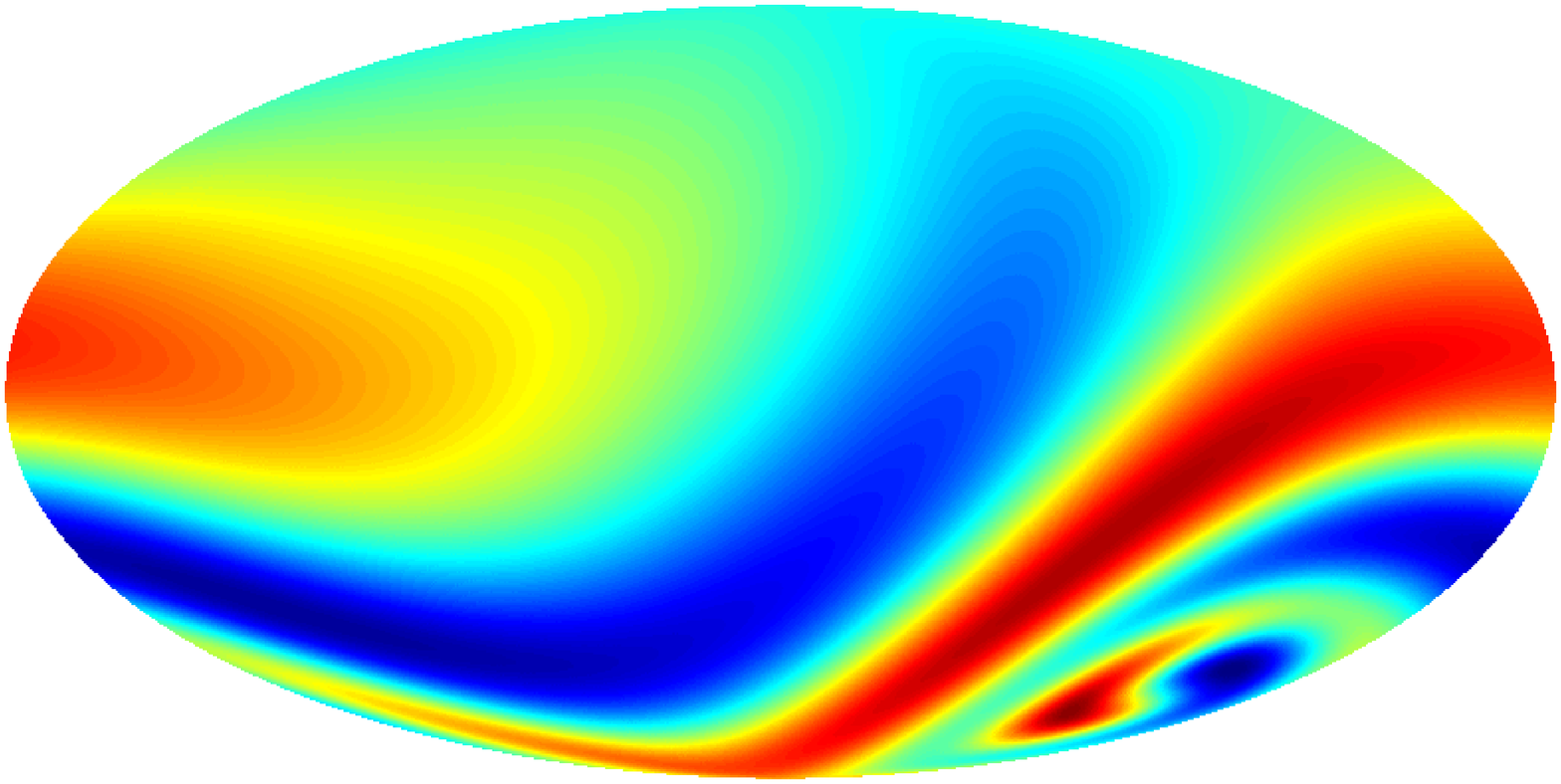}}
	\subfigure[Concordance cosmology]{
          \includegraphics[width=.5\columnwidth]{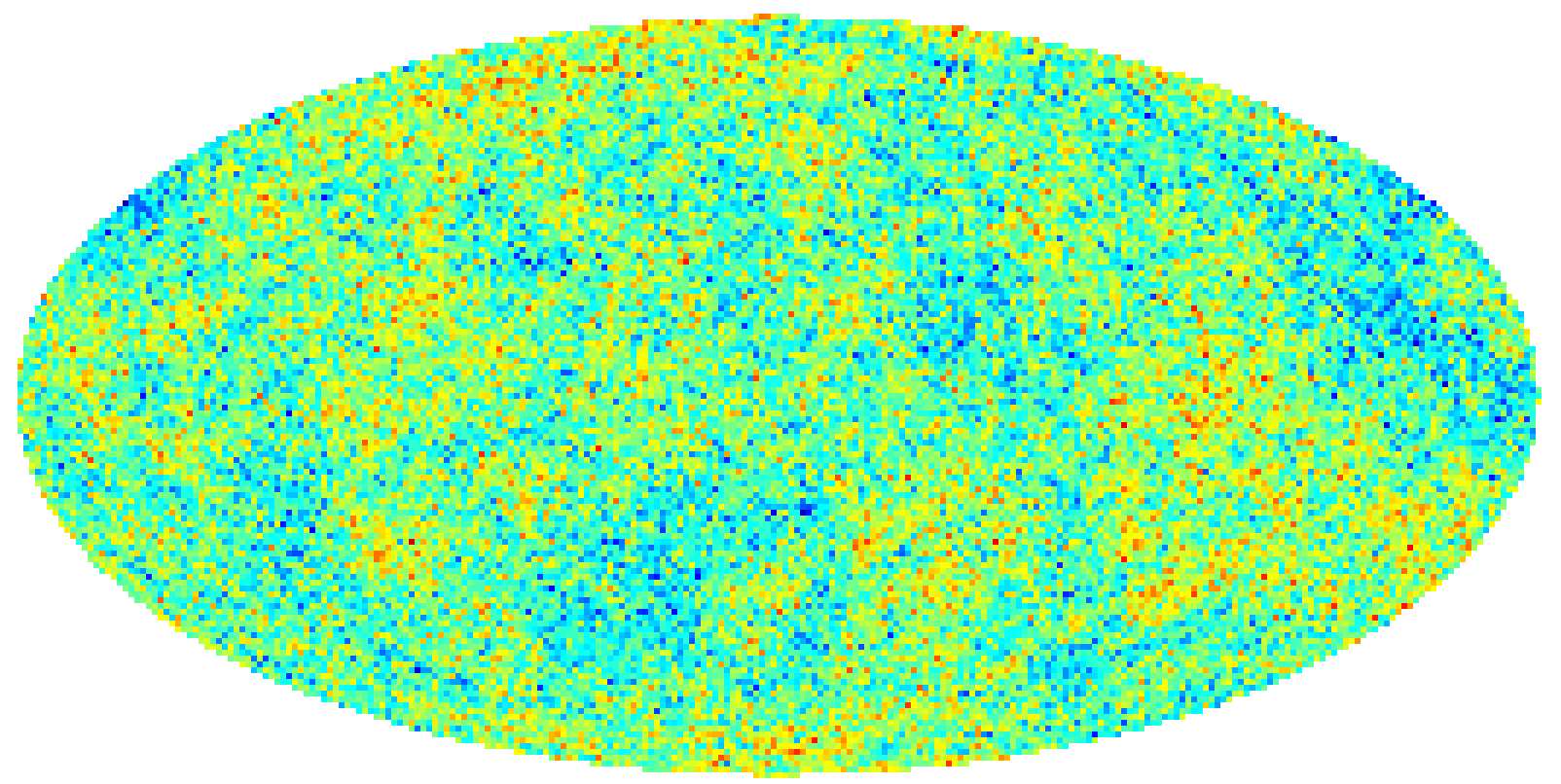}} \\
	\subfigure[$\omega = 5\times 10^{-10}$]{
          \includegraphics[width=.5\columnwidth]{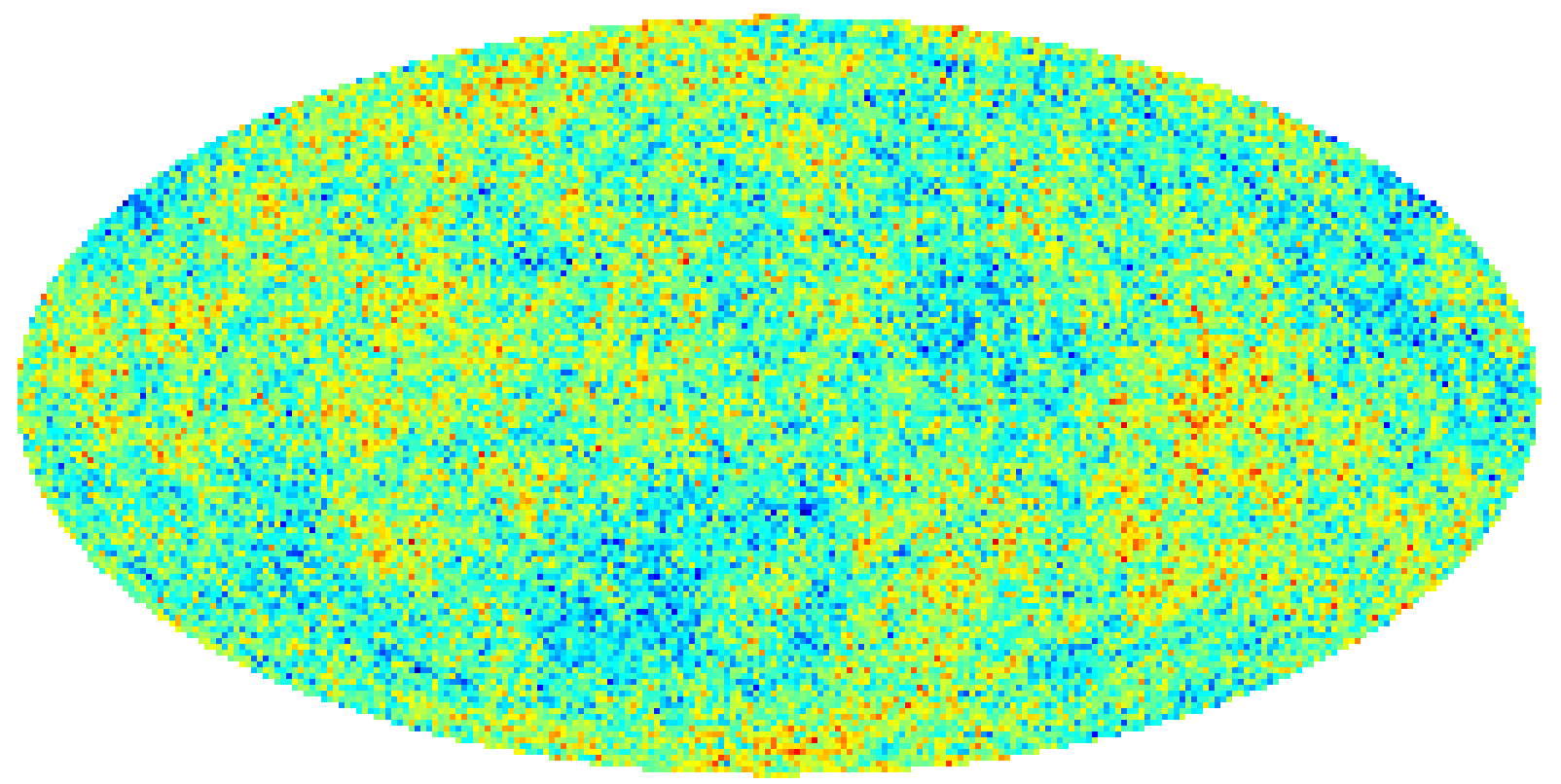}}
	\subfigure[$\omega = 7\times 10^{-10}$]{
          \includegraphics[width=.5\columnwidth]{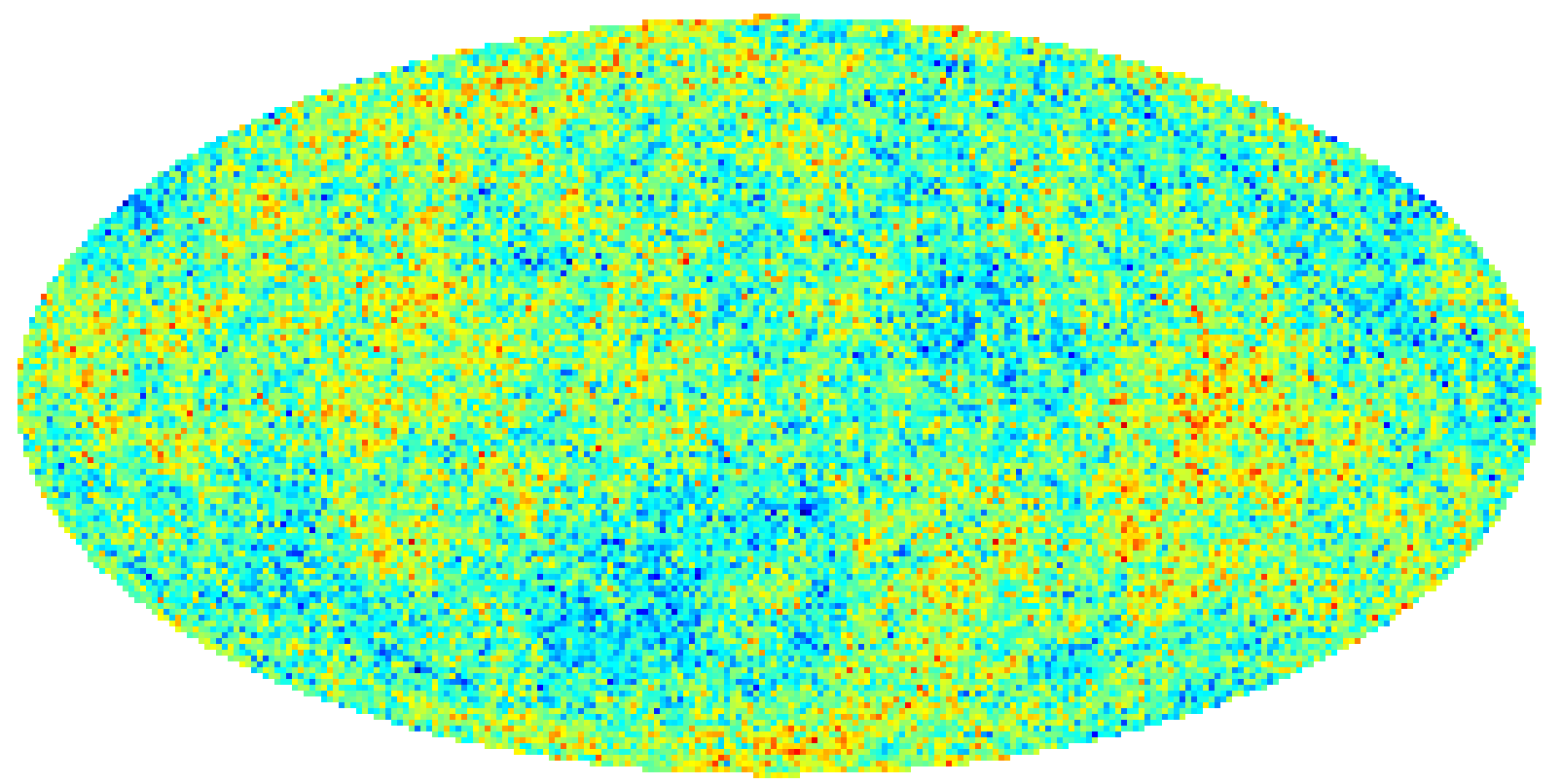}} \\
     	\subfigure[$\omega = 9\times 10^{-10}$]{
           \includegraphics[width=.5\columnwidth]{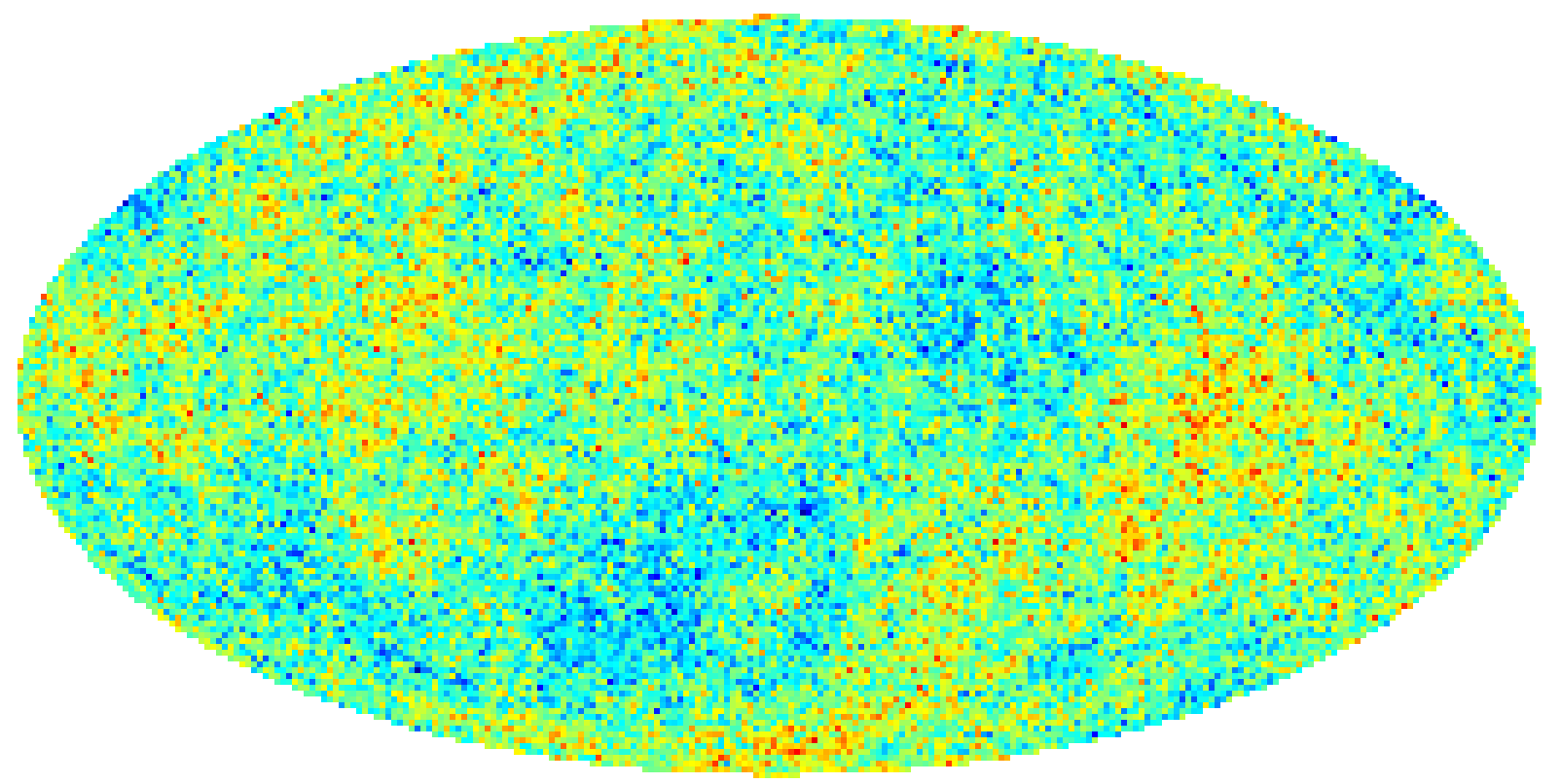}}
	\subfigure[$\omega = 11\times 10^{-10}$]{
           \includegraphics[width=.5\columnwidth]{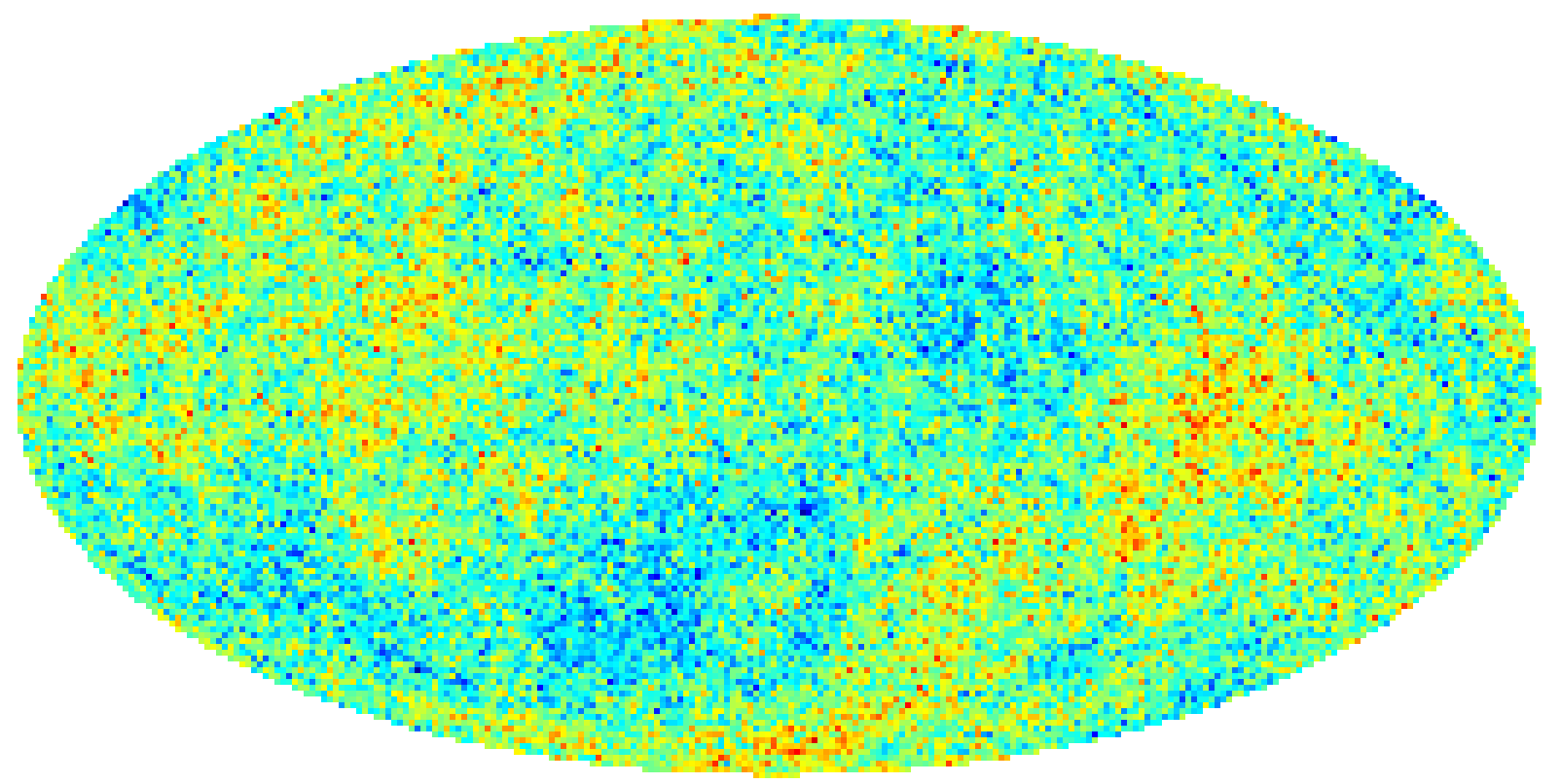}} \\
	\subfigure[$\omega = 13\times 10^{-10}$]{
           \includegraphics[width=.5\columnwidth]{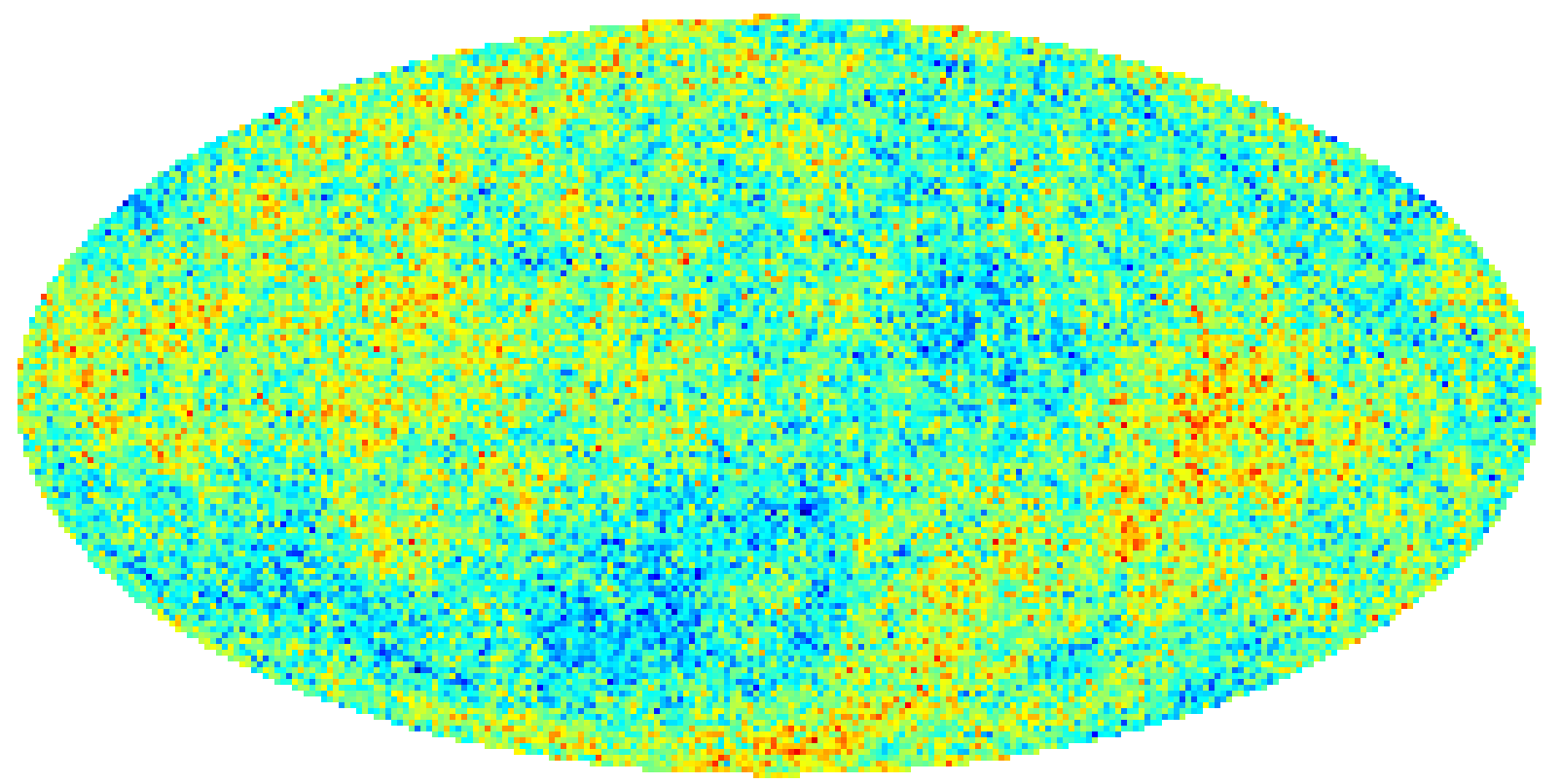}}
	\subfigure[$\omega = 15\times 10^{-10}$]{
           \includegraphics[width=.5\columnwidth]{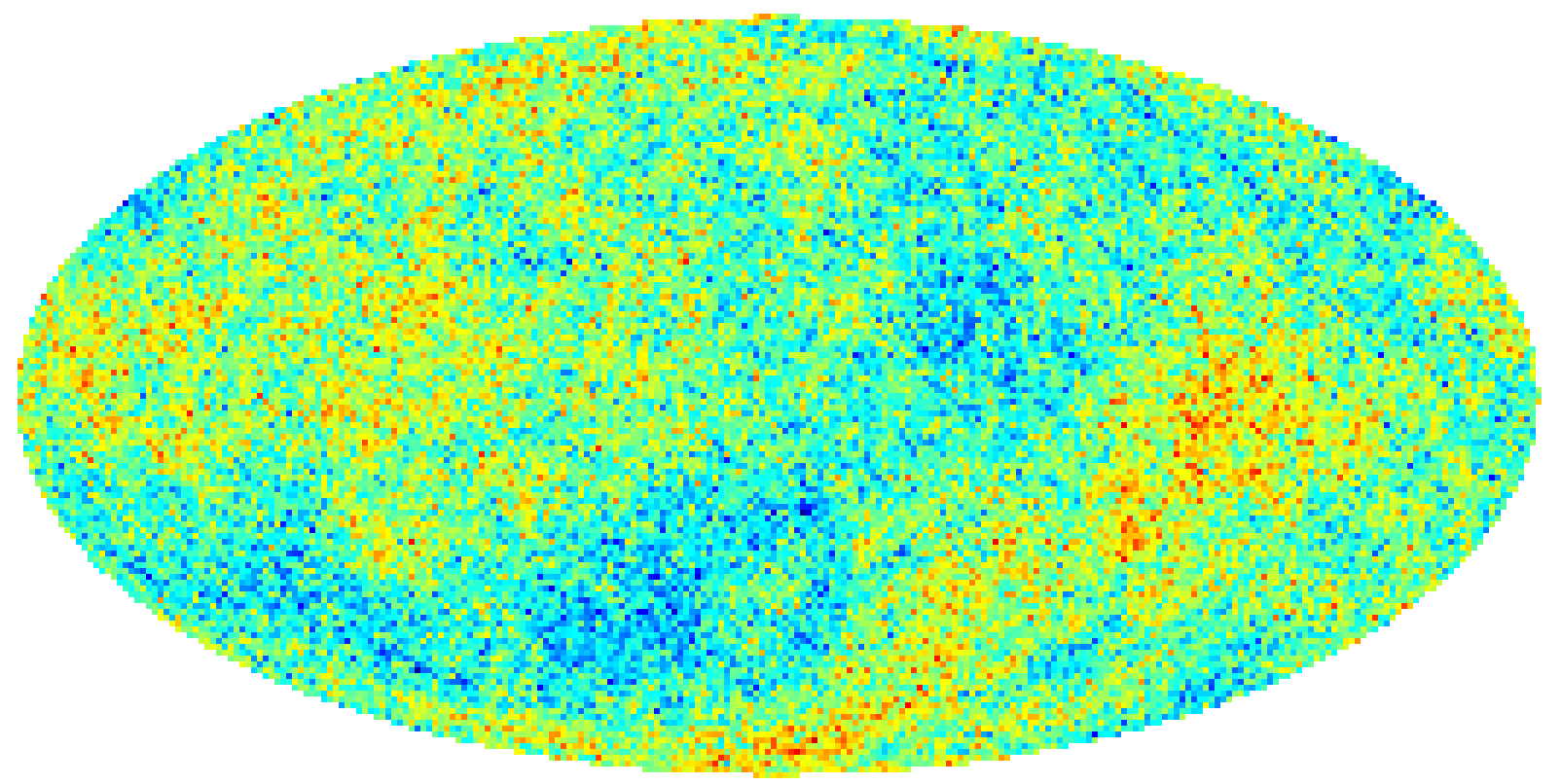}}\\
	 \subfigure[$\omega = 17\times 10^{-10}$]{
           \includegraphics[width=.5\columnwidth]{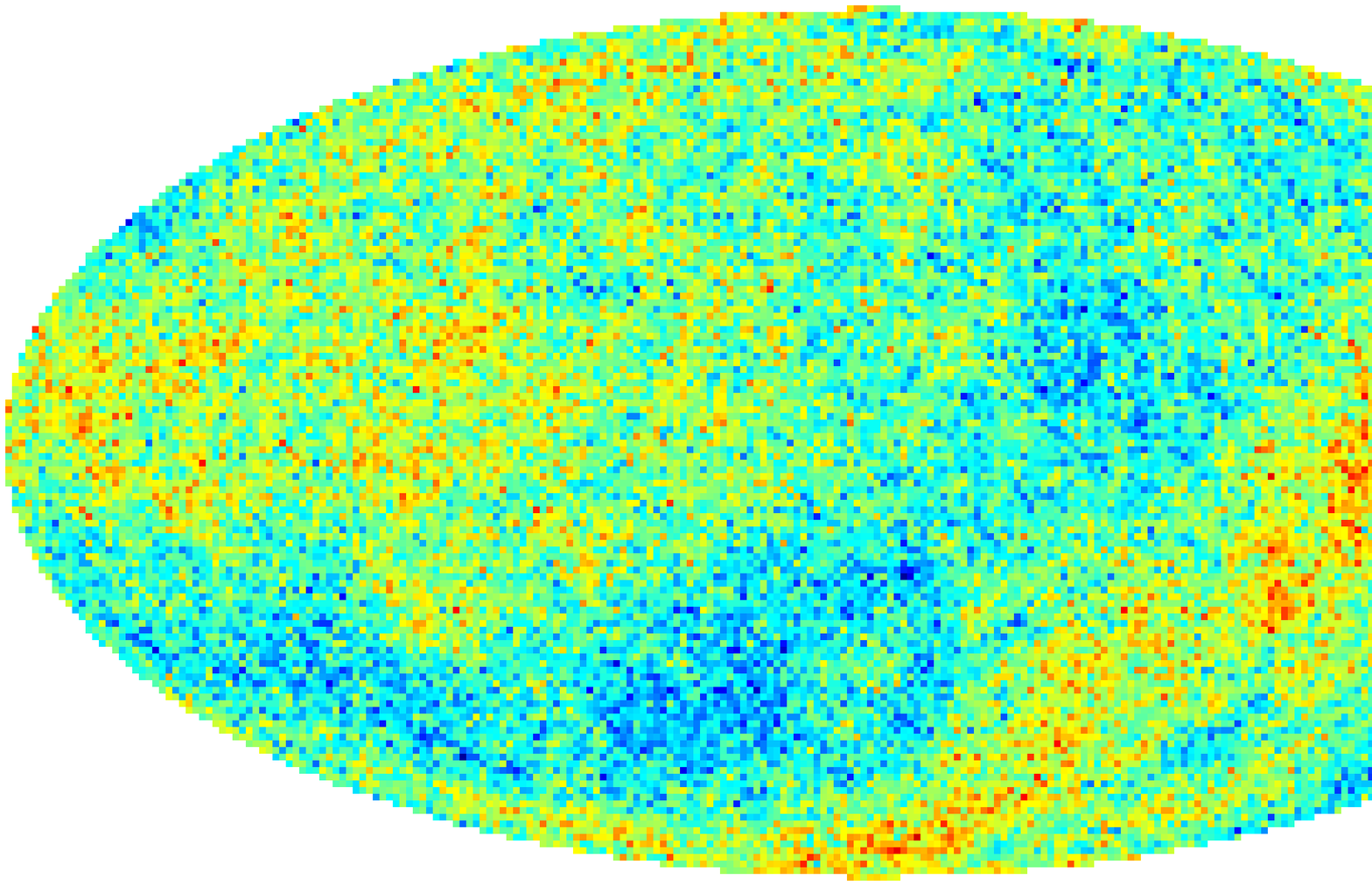}} 
	   \subfigure[$\omega = 19\times 10^{-10}$]{
           \includegraphics[width=.5\columnwidth]{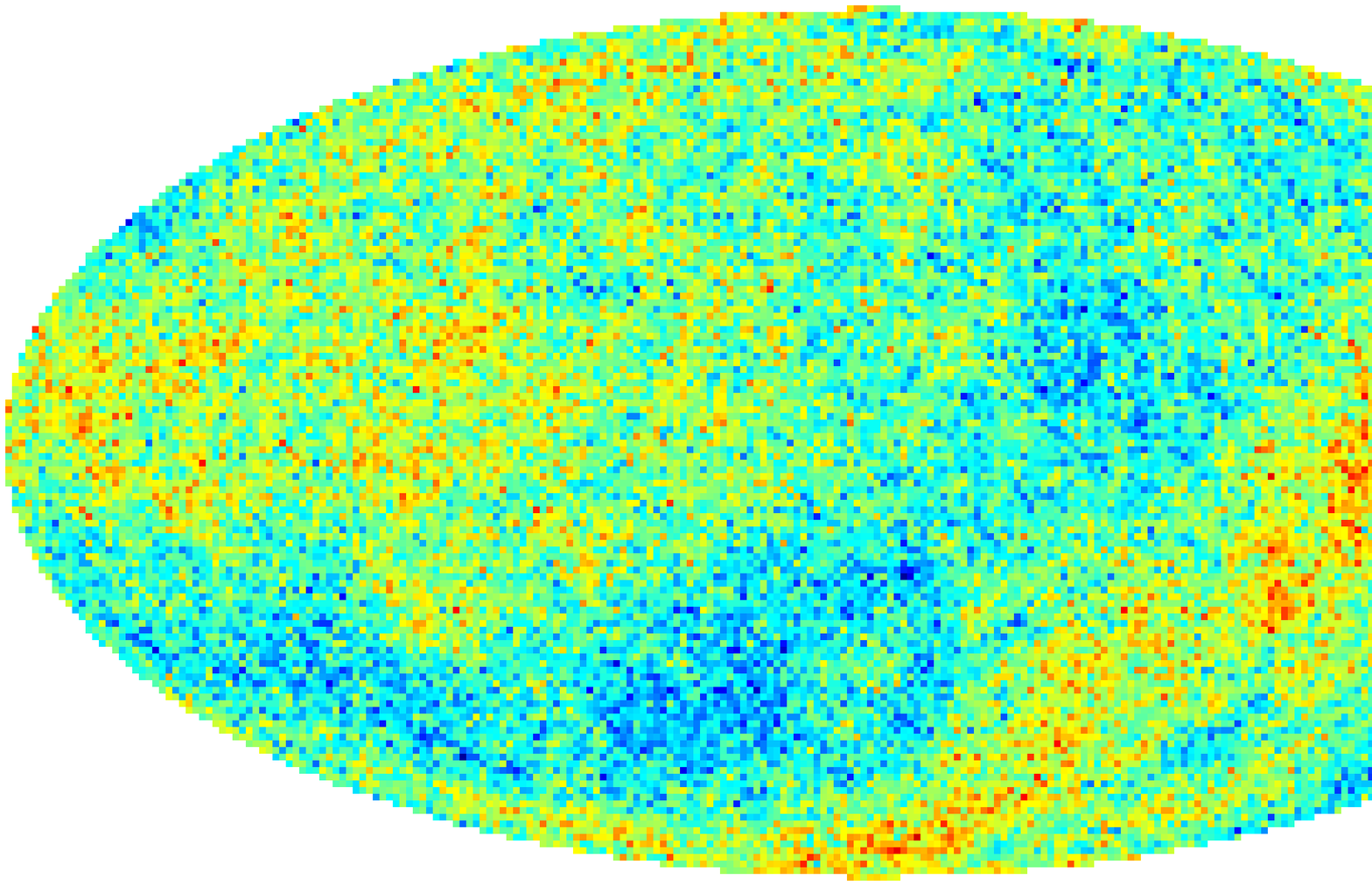}}\\
	   \begin{center}
	    \subfigure{
           \includegraphics[width=0.8\columnwidth]{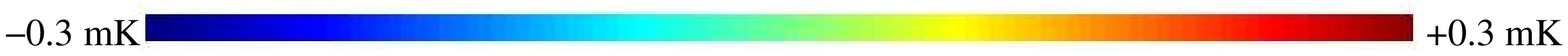}}\\
	   \end{center} 
	       	         
\caption{Simulated data created with a WMAP concordance cosmology (a) and a Bianchi component (enhanced by a factor of 10 for
clarity) (b) of the type found by Jaffe
 with an increasing vorticity ($\omega$).}
\label{simulations}
\end{figure}

In order to provide the best possible assessment of our method the Bianchi simulations were created with Jaffe's best fit parameter set of: $h=0.15$, $\Omega_m=0.5$, $\Omega_{\Lambda}= 0$, $\alpha = 42^{\circ}$, $\beta
= 28^{\circ}$ and $\gamma = 310^{\circ}$ (see Fig. \ref{simulations} (a)) combined with a best fit WMAP cosmology (see Fig.
\ref{simulations} (b)) with eight equally spaced vorticities incremented between $5$ and $19 \times 10^{-10}$ (see Fig.
\ref{simulations} (c)-(j)). Ultimately we aim to ascertain via these simulations the lowest amplitude of Bianchi signal we can succesfully
extract via our method --where $\omega = 10\times 10^{-10}$ represents the level of sensitivity required to detect a Jaffe type component in real data. 

Marginalised parameter constraints (see Fig. \ref{extracted}) illustrate we are able to successfully extract the input
Bianchi parameters for amplitudes at and above roughly $\omega = 11 \times 10^{-10}$, which suggests our method should be able to determine
parameters of any Bianchi component similar to the Jaffe template, should one exist. As for the question of whether such an inclusion is
actually warranted by the data, we consider the evidence. A useful guide has been given by \citet{Jeffreys} to rank relative evidence
differences in order of significance: $\Delta\mbox{ln} E < 1$ inconclusive, $1 < \Delta\mbox{ln} E < 2.5$ significant, $2.5 <
\Delta\mbox{ln} E < 5$ strong and $\Delta\mbox{ln} E > 5$ decisive. While there is marginal evidence in these simulations for a component
at $\omega = 11 \times 10^{-10}$ (see Table \ref{table1}) a significant model selection can only be made at $\omega = 13 \times 10^{-10}$,
also roughly at the point where the Bianchi template becomes visibly apparent in the sky maps in Fig. \ref{simulations}.

\begin{figure}
\begin{center}
\includegraphics[width=0.7\linewidth, angle = -90]{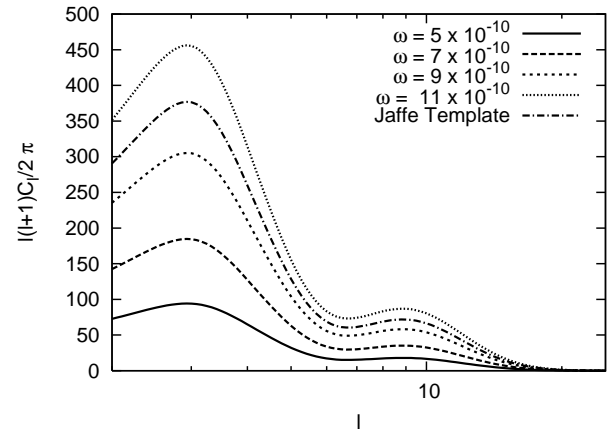}
\caption{Comparison of the relative amplitudes of the angular power spectra ($C_l$) of four simulated 
Bianchi skies of varying vorticity and the Jaffe template.}
\label{amplitude}
\end{center}
\end{figure}   

\begin{figure}
    	\subfigure{
	  \includegraphics[width=.3\columnwidth]{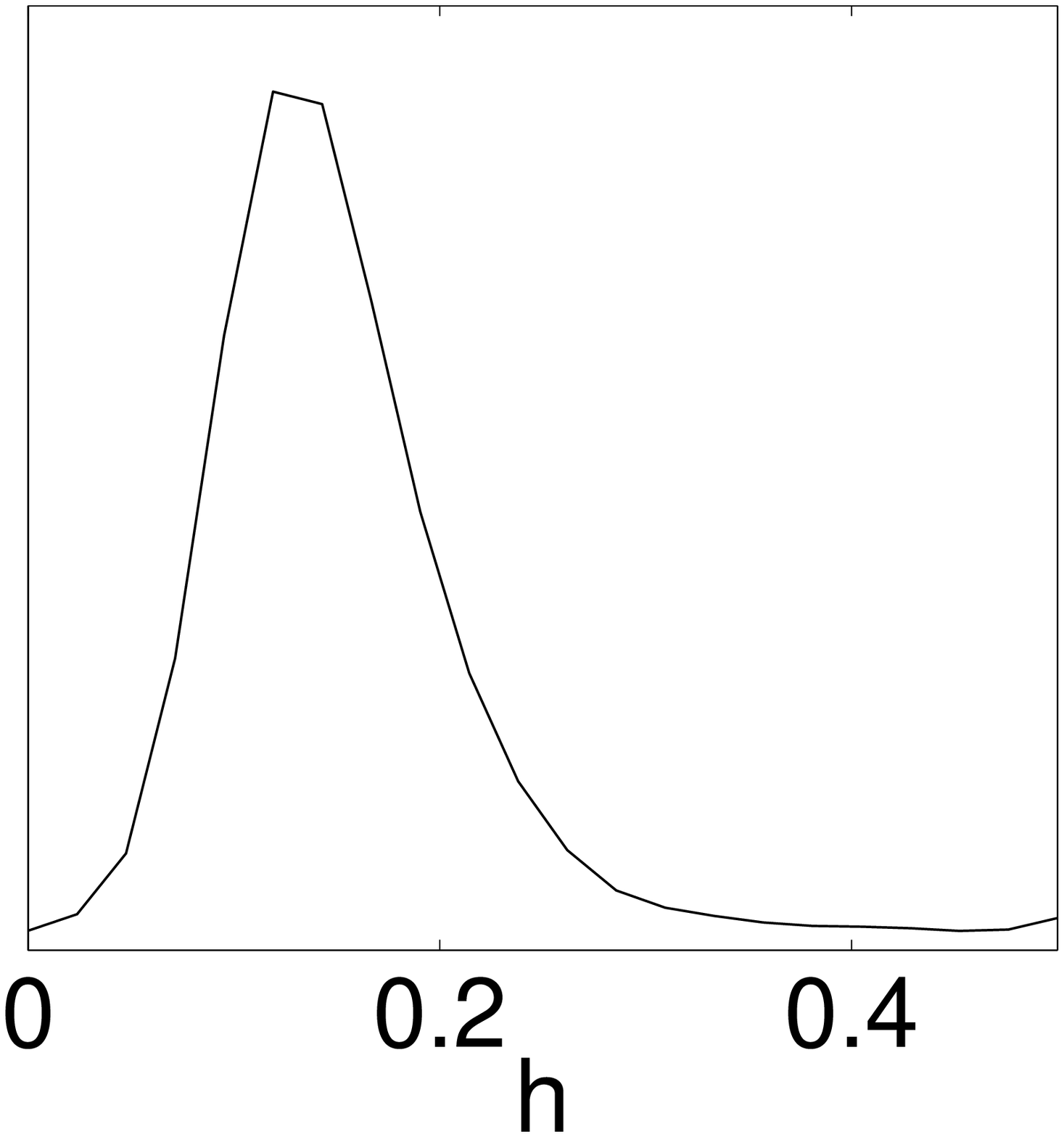}}
	  \hfill
	\subfigure{
	  \includegraphics[width=.3\columnwidth]{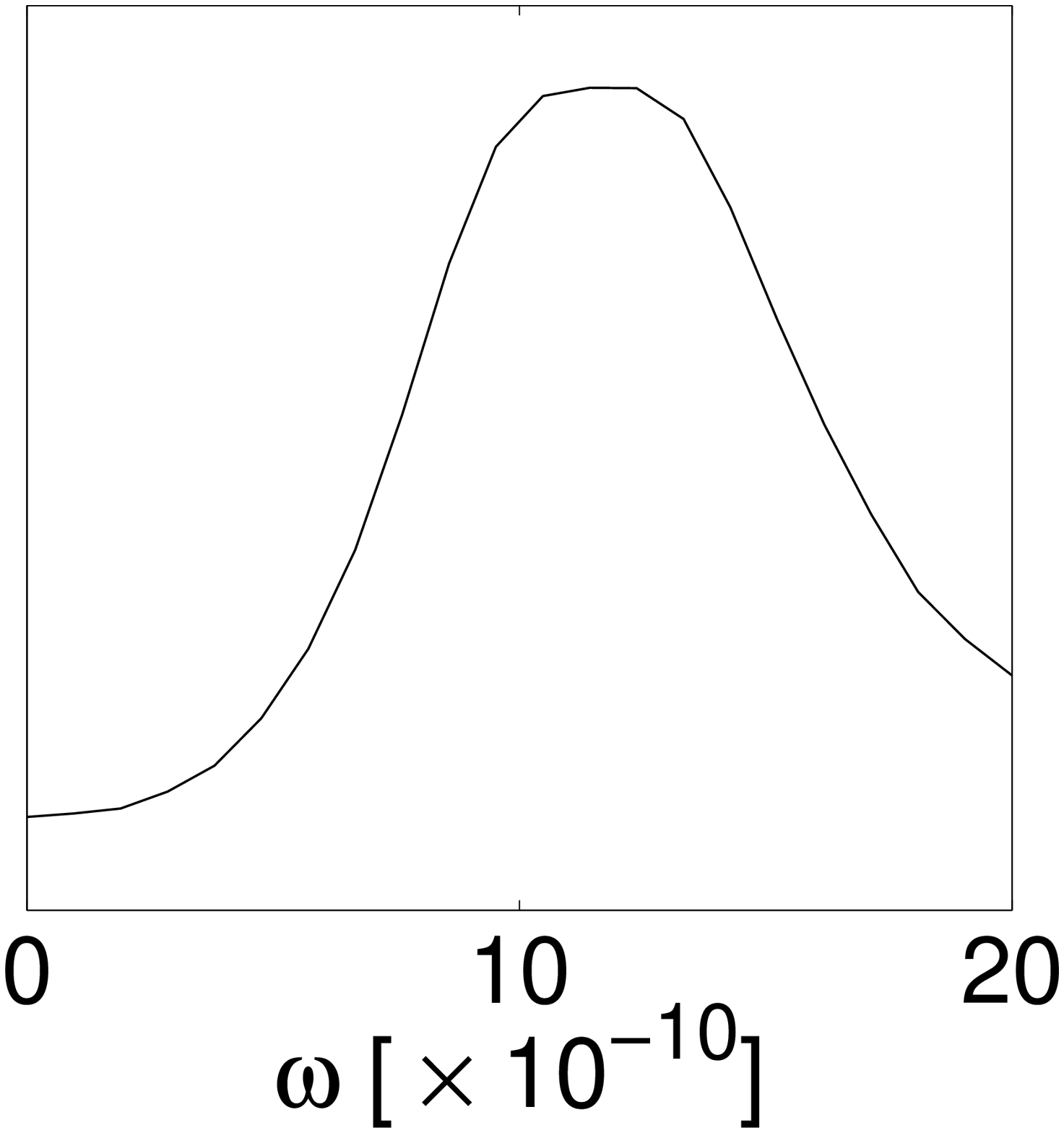}}
     	\hfill
	\subfigure{
	   \includegraphics[width=.31\columnwidth]{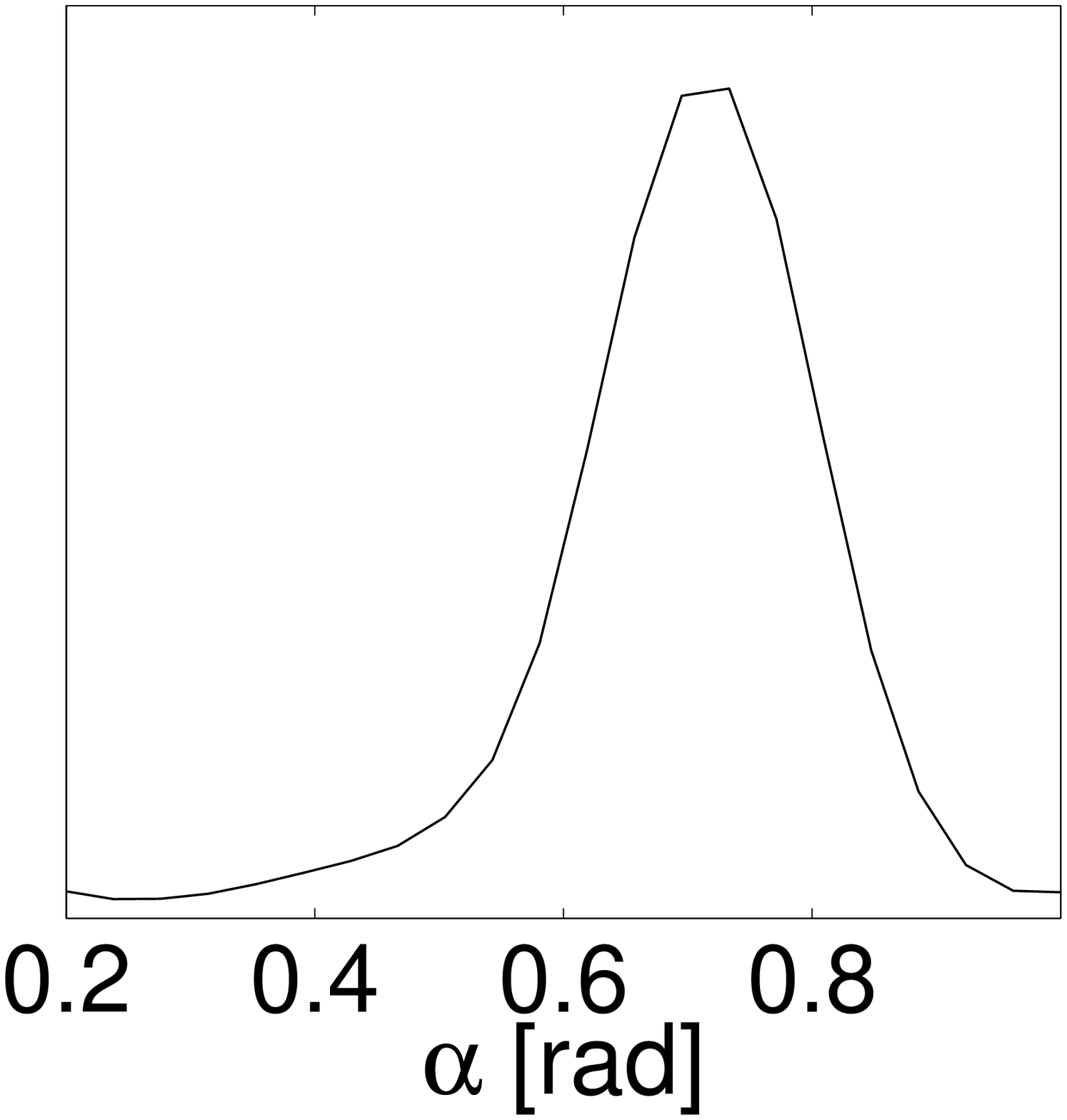}}\\
	\subfigure{
	   \includegraphics[width=.305\columnwidth]{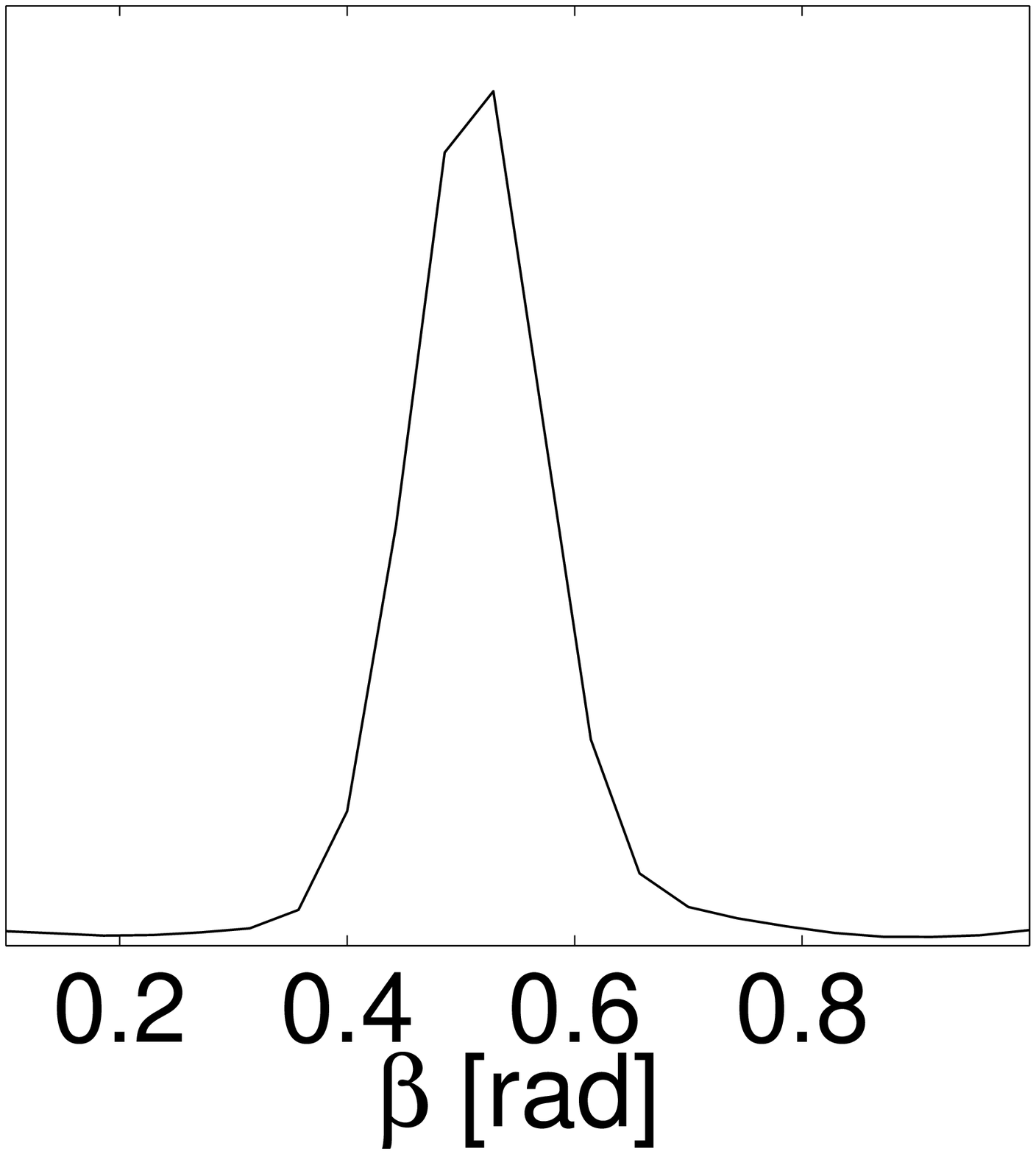}}
	\hfill
	\subfigure{
	   \includegraphics[width=.3\columnwidth]{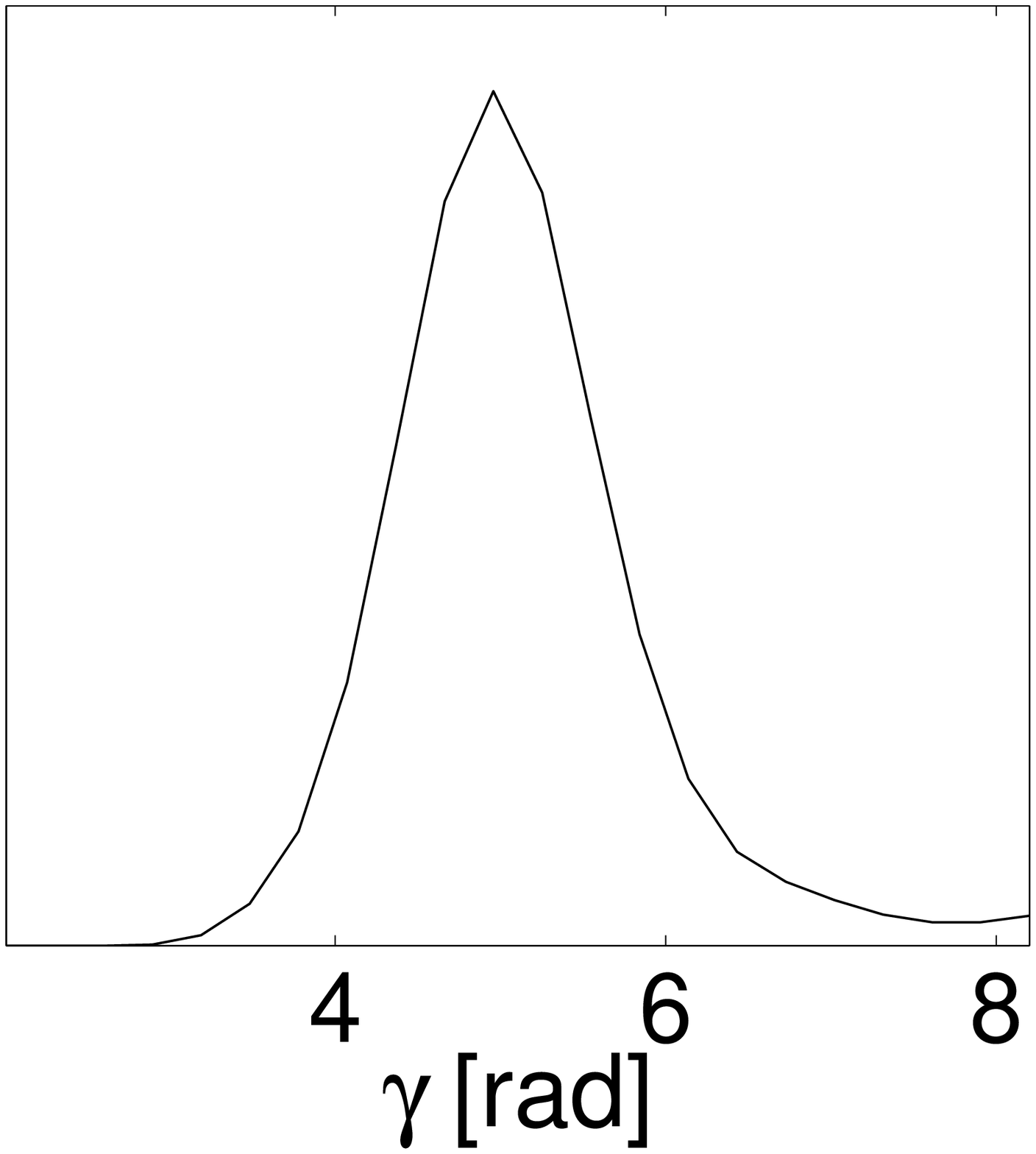}}
	\hfill
	\subfigure{
	   \includegraphics[width=.3\columnwidth]{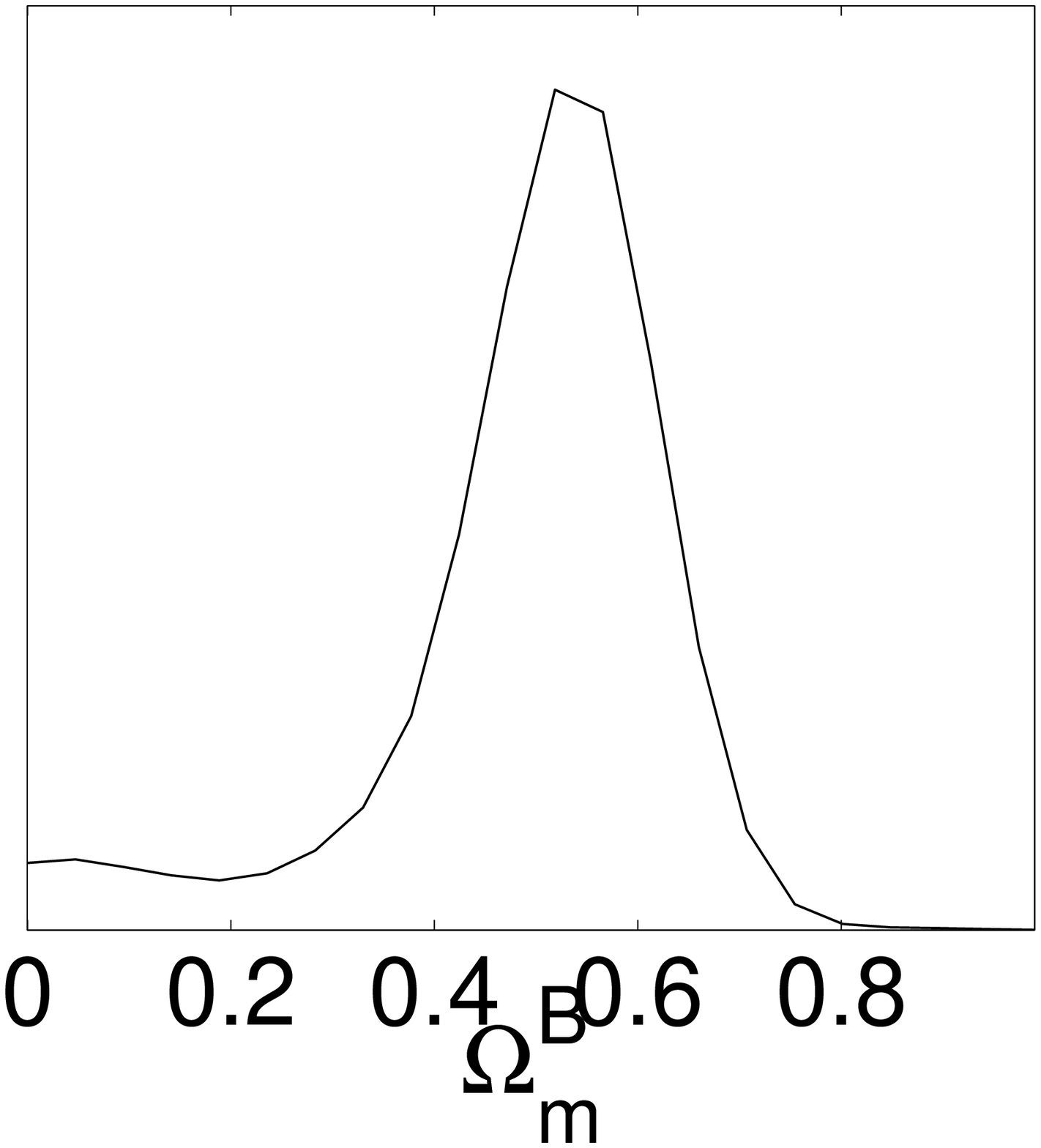}}\\
	\hfill
	  \subfigure{
           \includegraphics[width=.3\columnwidth]{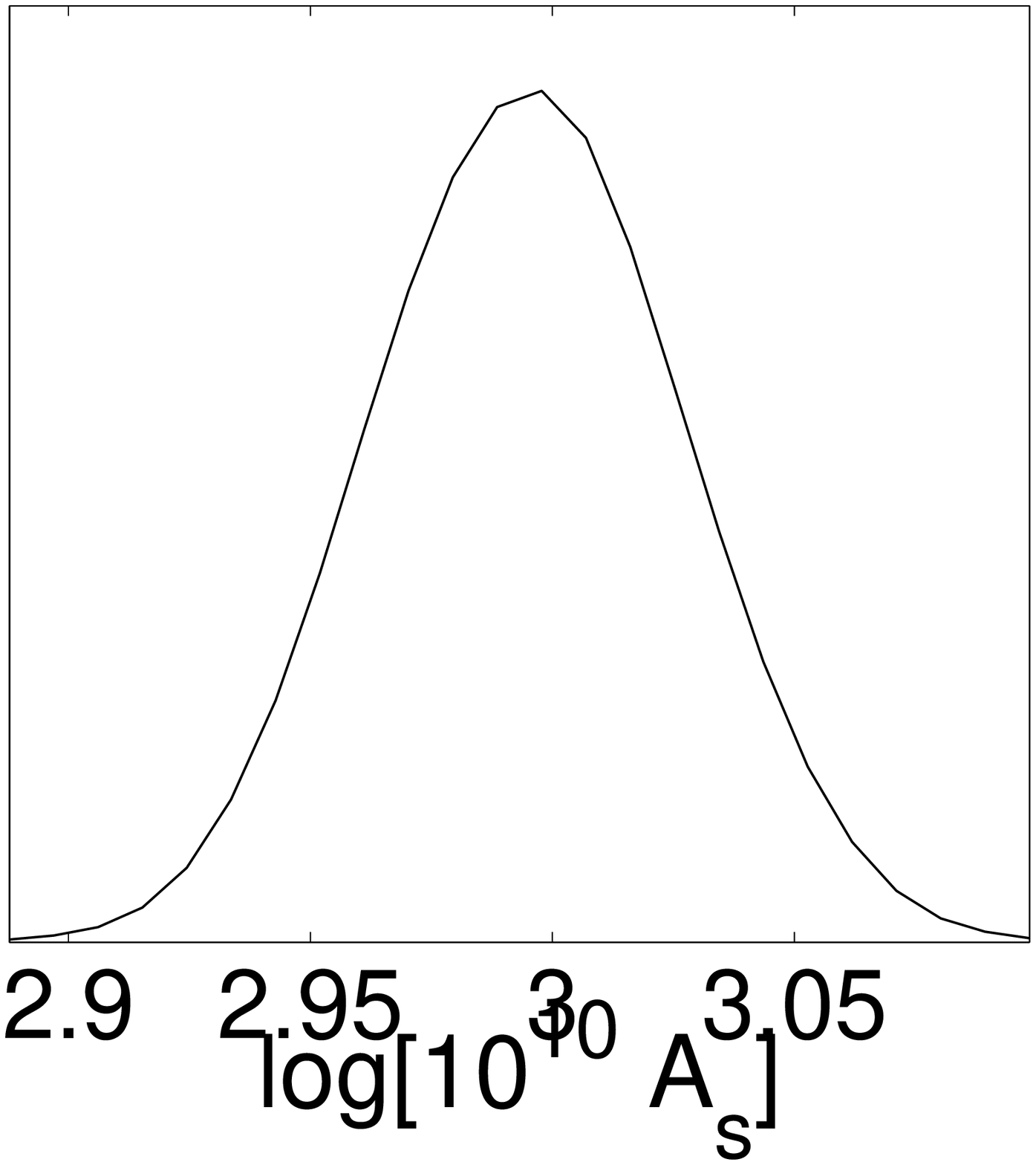}}
\caption{Extracted cosmological and Bianchi parameters from a simulated map with $\omega = 11 \times 10^{-10}$
(described in text).}
\label{extracted}
\end{figure}

\begin{table}
\begin{center}
\caption{Differences of $\ln$ evidence with (model A) and without (model B) a Bianchi component with vorticity incremented between
$5-19\times 10^{-10}$.}
\begin{tabular}{|c||c||c|}
    \hline
 $\mathbf{\omega}$ $[\times 10^{-10}]$&  \textbf{A} & \textbf{B} \\
    \hline
 5 &0.0 $\pm 0.2$ &  -1.6 $\pm$ 0.3\\  
 7 &0.0 $\pm 0.2$ &  -1.3 $\pm$ 0.3\\
 9 &0.0 $\pm 0.2$ &  -1.1  $\pm$  0.3 \\
 11 &0.0 $\pm 0.2$ & -0.9 $\pm$ 0.3\\
 13 &0.0 $\pm 0.2$ & +1.4 $\pm$ 0.3\\
 15 &0.0 $\pm 0.2$ & +3.6 $\pm$ 0.3\\
 17 &0.0 $\pm 0.2$ & +6.2  $\pm$ 0.3\\
 19&0.0 $\pm 0.2$ &  +8.4 $\pm$ 0.3\\
    \hline
\end{tabular}
\label{table1}
\end{center}
\end{table}

\section{Application to Real Data} \label{Real Data}
The analysis was performed on both one and three year ILC maps, probing structure
up to multipoles of 64, which does not limit any study of Bianchi structure, which
contains few features beyond $l = 20$ (see Fig. \ref{amplitude}) but does preclude
a full examination of the background cosmology. This was an unfortunate necessity
of using the ILC maps owing to their complex noise properties. 

As we are adopting a Bayesian approach in this analysis our results will be
dependent on the prior parameter ranges chosen. Our most conservative, and thus 
widest uniform prior ranges are listed in Table \ref{priors}. From a Bayesian point
of view these are broad enough to represent the situation in which there is little prior knowledge of
the Bianchi models. In particular the $h$ parameter,
which decreases the spiral `tightness' almost exponentially between 0.01 and 0.2
but little thereafter was given a prior range from 0.01 to 1 --encompassing the
majority of possible features. The vorticity $\omega$ increases the Bianchi
amplitude roughly linearly so a prior range centred on the amplitude of the Jaffe
template was adopted providing a realistic upper limit of $20 \times 10^{-10}$ at
which point the Bianchi signal would dominate the CMB sky and a lower limit of
$\omega = 0$ corresponding to no Bianchi signal. Bianchi ${\rm VII_h}$ models are
only defined in open cosmologies where $\Omega_{tot} < 1$\footnote{Similarly
Bianchi ${\rm VII_0}$ and ${\rm IX }$ are only defined in flat and closed
cosmologies respectively.} so the prior chosen on the combination of matter and
dark energy energy density was restricted to lie within $0.01$ and $0.99$ --thus
introducing an implicit prior on $\Omega_{\Lambda}^{\rm B}$ of $\pm 0.8$ (as
$\Omega_{\Lambda}^{\rm B} + \Omega_m^{\rm B} = \Omega_{tot}^{\rm B}$). The Euler angles position the centre
of the spiral, via $\alpha$ and $\beta$ varied over $360^{\circ}$ and $180^{\circ}$
respectively while the orientation of the pattern is given by $\gamma$ rotating
over $360^{\circ}$. In the subsequent sections we will analyse a range of Bianchi
models (Table \ref{models}) defined by subsets of the above prior ranges while
simultaneously fitting a scalar perturbation amplitude $A_s$ of an assumed
background concordance cosmology. In Section \ref{Model Selection} we will compare,
via the evidence, models B -- G with the concordance cosmology model A as to their
ability to describe, adequately the observed CMB sky.    

\begin{table}
\begin{center}
\caption{Summary of Bianchi ${\rm VII_h}$ component parameters and priors.}
\begin{tabular}{|c|}
    \hline
 \textbf{Full Bianchi}\\
    \hline
 $\Omega_{tot}^{\rm B} = [0.01,0.99]$\\  
 $\Omega_{m}^{\rm B} = [0.01,0.99]$\\
 $h = [ 0.01,1]$\\
 $\omega = [0 ,20 ]\times 10^{-10}$\\
 $\alpha = [0 , 2\pi]$${\rm rads}$\\
 $\beta = [ 0, \pi]$${\rm rads}$\\
 $\gamma = [ 0, 2\pi ]$${\rm rads}$\\
 Chirality = L/R\\
    \hline
\end{tabular}
\label{priors}
\end{center}
\end{table}

\begin{table}
\begin{center}
\caption{Cosmological and Bianchi parameterisations for each of the parameter subsets studied.}
\begin{tabular}{|c||c||c|}
    \hline
 \textbf{Model} &  \textbf{Cosmology} & \textbf{Bianchi}\\
    \hline
 A & $A_s$ & -\\
 B & $A_s$ & $\Omega_m^B$, $\Omega_{tot}^B$, $h$, $\omega$, $\alpha$, $\beta$, $\gamma$, L/R\\
 C & $A_s$ & $\Omega_m^B$, $\Omega_{tot}^B$, $h$, $\omega$, $\alpha$--[0.4,1], $\beta$--[0.1, 0.7], $\gamma$, L/R\\
 D & $A_s$ & $\Omega_m^B$, $\Omega_{tot}^B$, $h$, $\omega$, $\gamma$, L/R\\
 E & $A_s$ & $\Omega_m^B$, $h$, $\omega$, $\alpha$, $\beta$, $\gamma$, L/R\\
 F & $A_s$ & $\Omega_m^B$, $h$, $\omega$, $\alpha$--[0.4,1], $\beta$--[0.1, 0.7], $\gamma$, L/R\\
 G & $A_s$ & $\Omega_m^B$, $h$, $\omega$, $\gamma$, L/R\\
 \hline 
\end{tabular}
\label{models}
\end{center}
\end{table}

\subsection{Parameter Constraints}  \label{Parameter constraints}
\begin{figure}
\includegraphics[width=\linewidth]{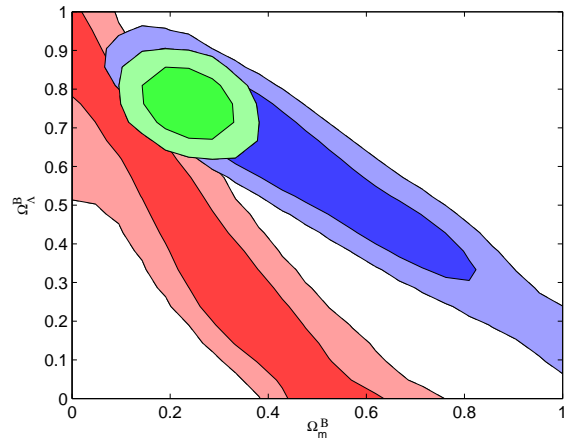} 
\caption{Comparison
of the $\Omega_m^{\rm B} - \Omega_{\Lambda}^{\rm B}$ Bianchi degeneracy (shaded with 1 and 2$\sigma$
contours) with the familiar CMB geometric degeneracy from WMAP first (blue) and third year + polarisation (green) data (with 1 and 2$\sigma$
contours).} 
\label{degeneracy}  
\end{figure}

The findings of \citet{Jaffea}, \citet{Jaffeb} and \citet{Jaffec} all point to a left-handed
Bianchi ${\rm VII_h}$ component with a total energy density $\Omega_{tot}^{\rm B} =
0.5$, this value being required to focus the centre of the spiral, thus alleviating, at
least partially, the observable non-Gaussianity (\citealt{Jaffea}; \citealt{McEwena}; \citealt{McEwenb})
if centered in the southern hemisphere. However both 1-year and 3-year WMAP data would
suggest a value of $\Omega_{tot}$ much closer to unity, perhaps larger, motivating the
search for a Bianchi model within a more viable cosmology. The extension of the Bianchi
parameter space to include a dark energy density $\Omega_{\Lambda}^{\rm B}$ should thus
be explored thoroughly to alleviate this tension. \citet{Jaffeb} found that a degeneracy
is introduced with $\Omega_{\Lambda}^{\rm B}$ in the $\Omega_m^{\rm B} -
\Omega_{\Lambda}^{\rm B}$ plane, similar to the `geometric' degeneracy which exists in
the CMB. Along this region, Bianchi skies appear identical, only being distinguished by
their relative amplitudes.  The maximum ln-likelihood of a Bianchi component at
$\Omega_m^{\rm B} = 0.5$, $\Omega_{\Lambda}^{\rm B} = 0$ is -2716.31. At locations that are broadly
consistent with concordance, such as $\Omega_m^{\rm B} = 0.3$, $\Omega_{\Lambda}^{\rm B} = 0.69$ or
$\Omega_m^{\rm B} = 0.2$, $\Omega_{\Lambda}^{\rm B} = 0.79$ this falls to -2717.81 and -2719.11
respectively.  A more thorough exploration of the space by MCMC sampling (Fig.
\ref{degeneracy}.) confirms that despite the increased degree of freedom that
$\Omega_{\Lambda}^{\rm B}$ introduces, at the $1 \sigma$ level no overlap with either 1 or
3-year WMAP likelihood contours is found. Based on these results we can effectively rule
out Bianchi ${\rm VII_h}$ models as the physical origin of this effect. Of course the
corrections that a Bianchi component makes to the WMAP data are no less interesting,
whether we can model them or not. This is especially true given that the areas
of non-Gaussianity have been mostly preserved into the 3-year WMAP release
\citep{McEwenb}.

\begin{figure}
    	\subfigure{
	  \includegraphics[width=.3\columnwidth]{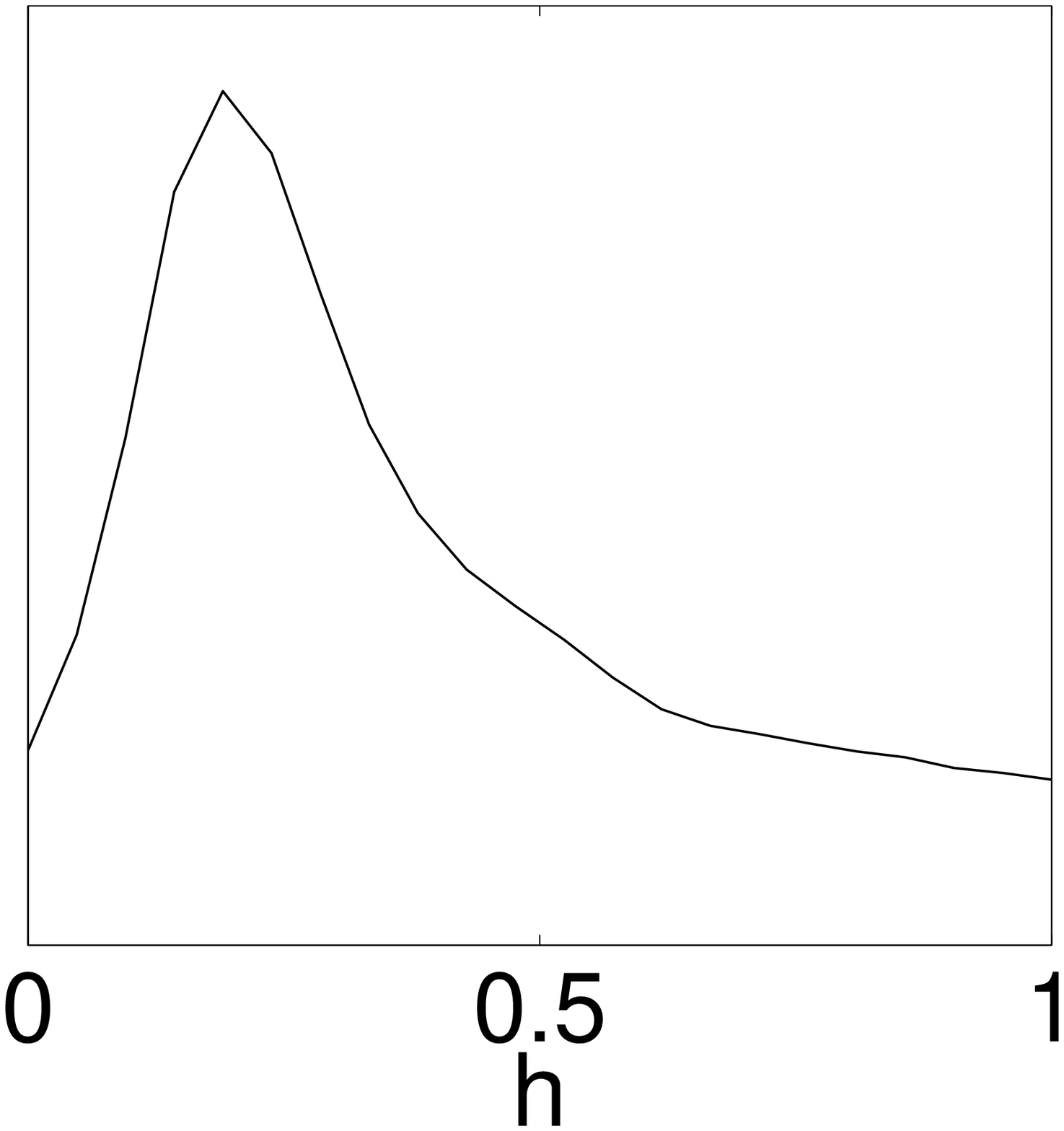}}
	  \hfill
	\subfigure{
	  \includegraphics[width=.3\columnwidth]{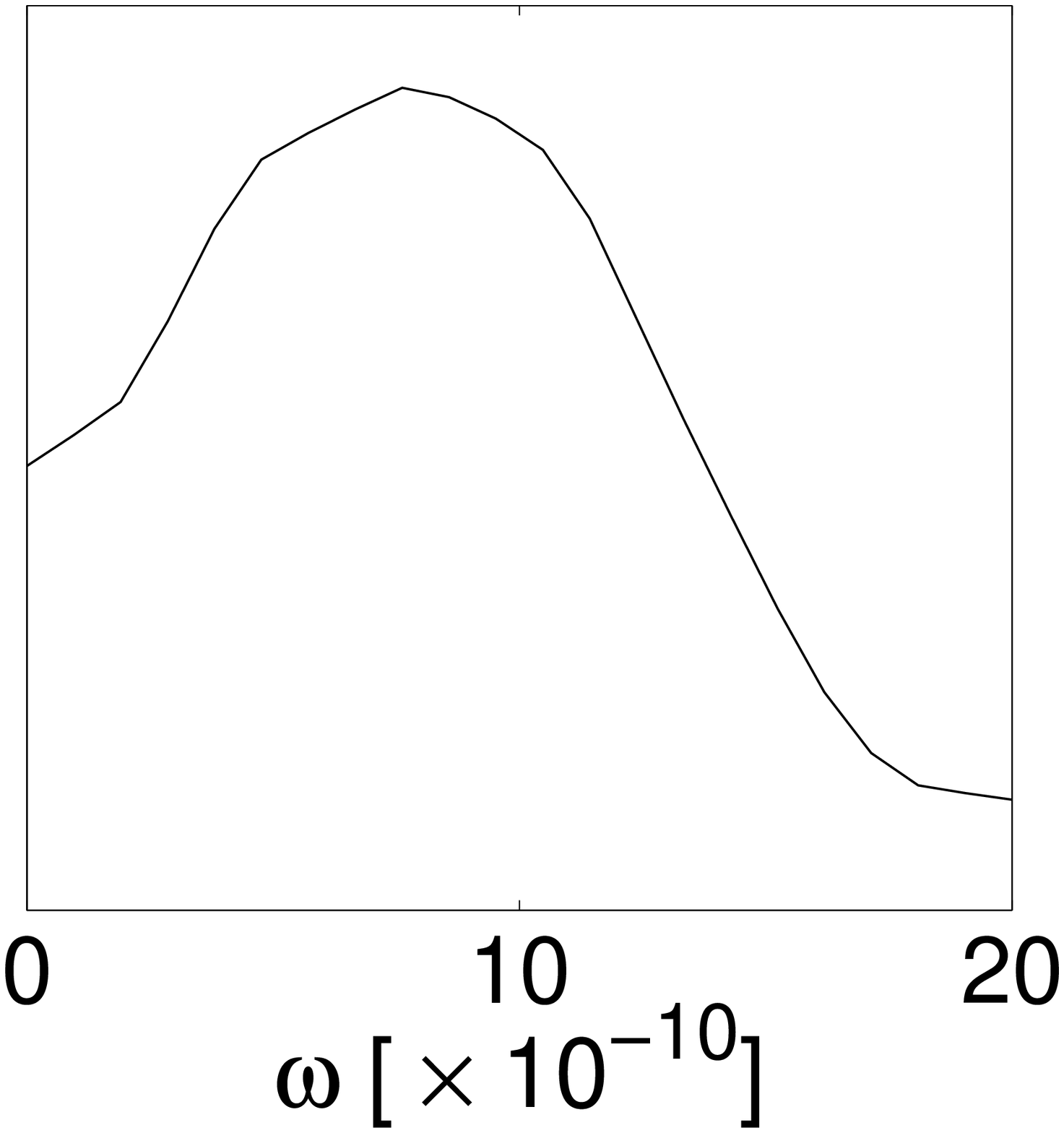}}
     	\hfill
	\subfigure{
	   \includegraphics[width=.31\columnwidth]{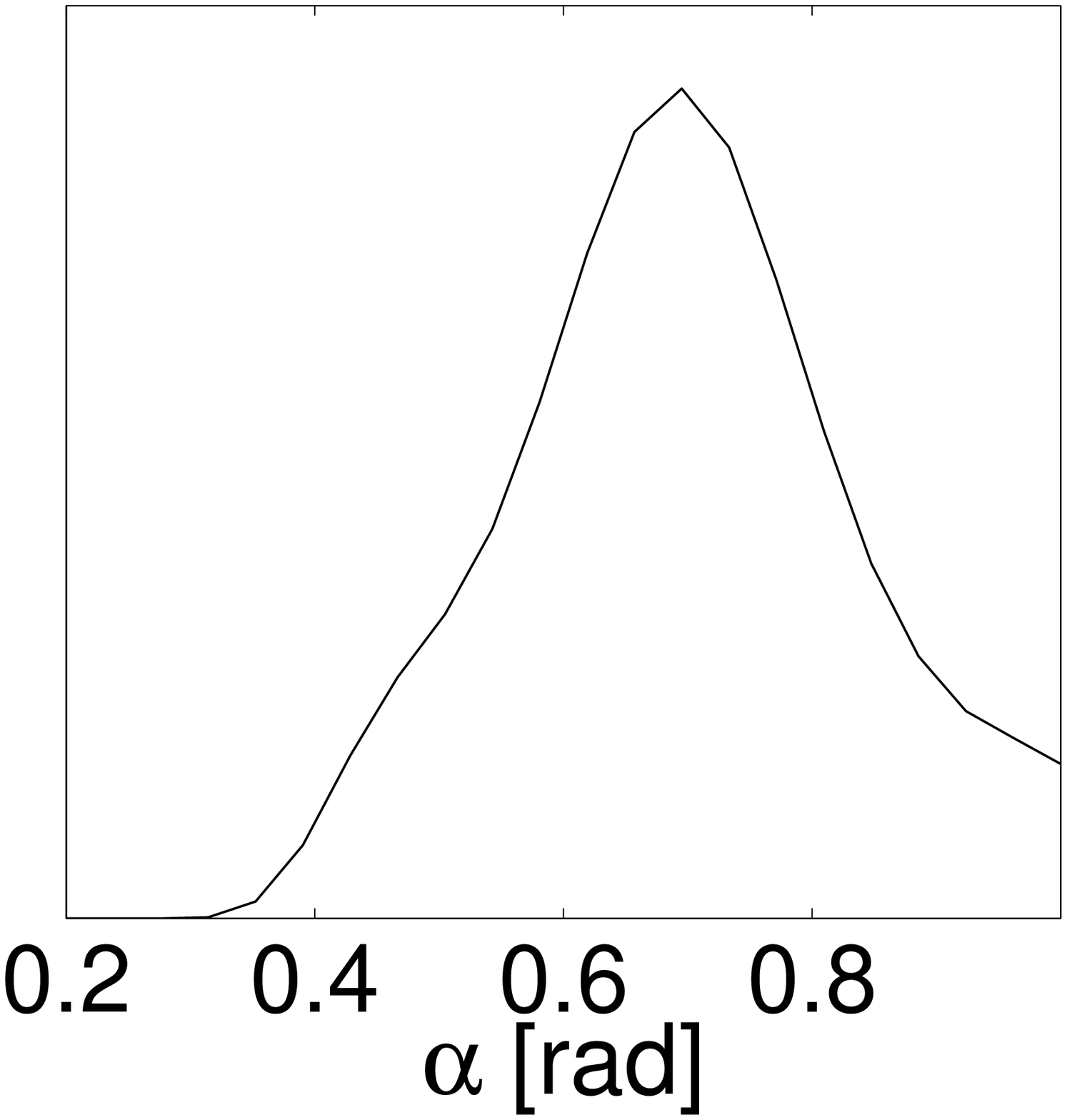}}\\
	\subfigure{
	   \includegraphics[width=.305\columnwidth]{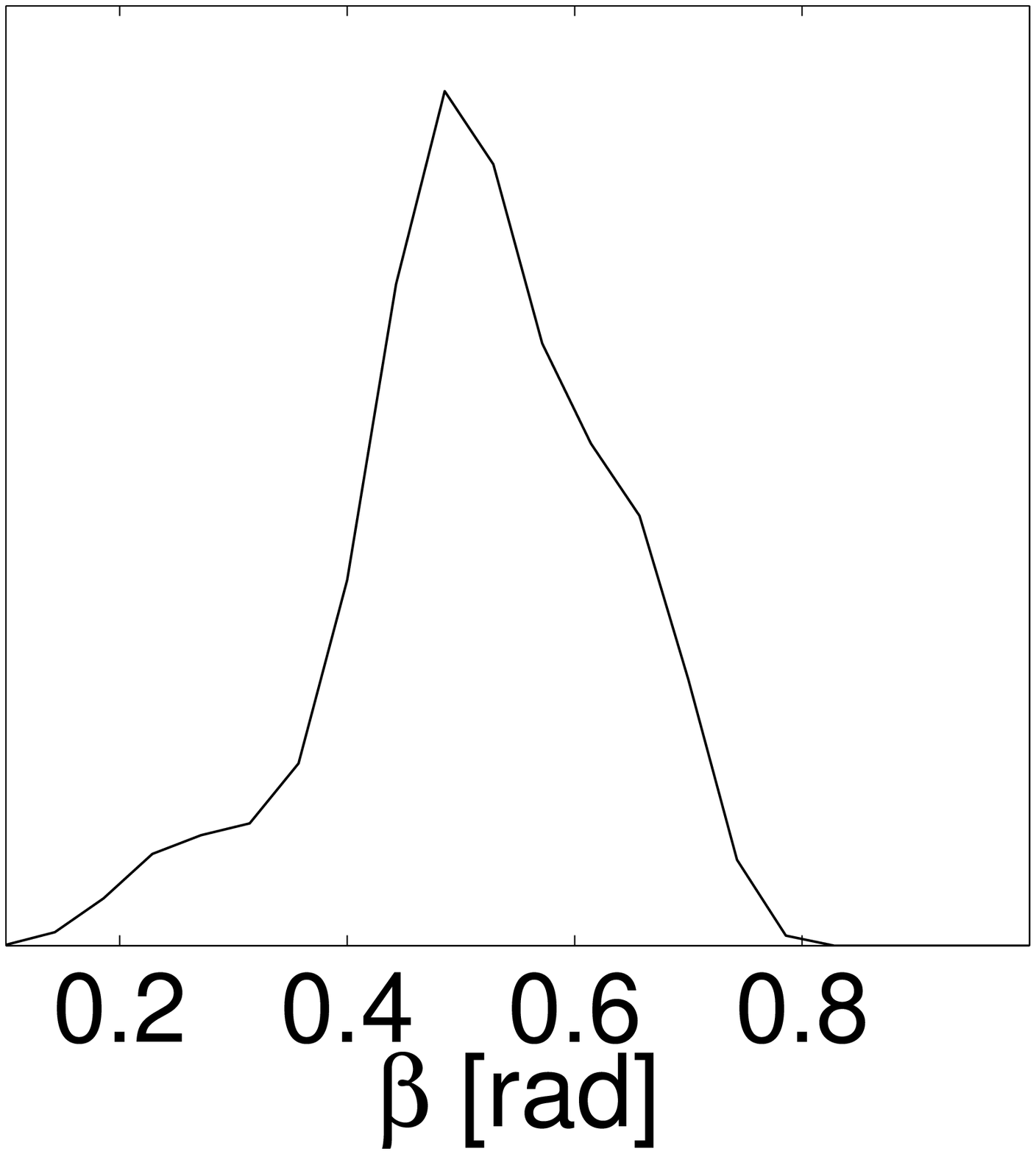}}
	\hfill
	\subfigure{
	   \includegraphics[width=.3\columnwidth]{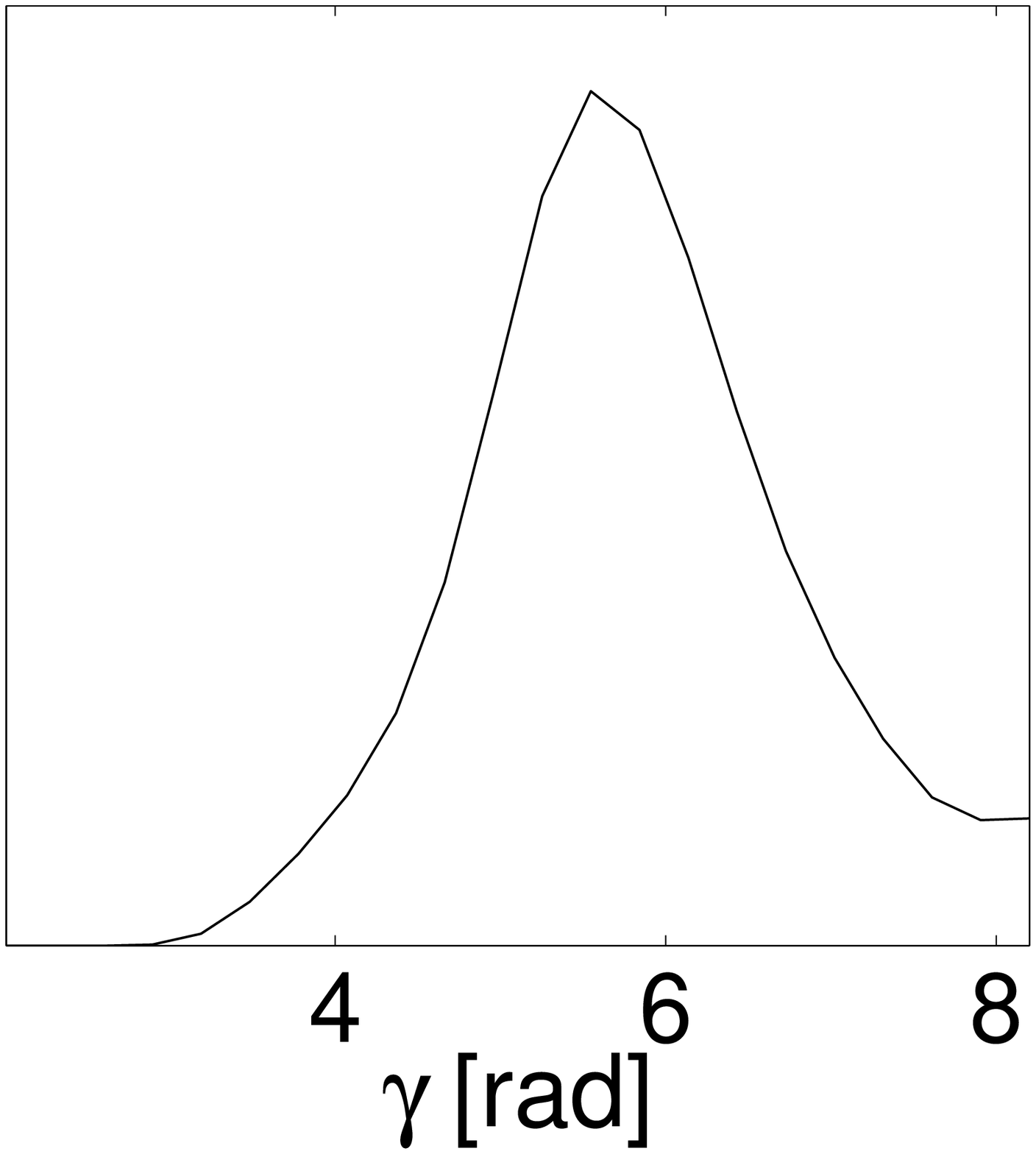}}
	\hfill
	\subfigure{
	   \includegraphics[width=.3\columnwidth]{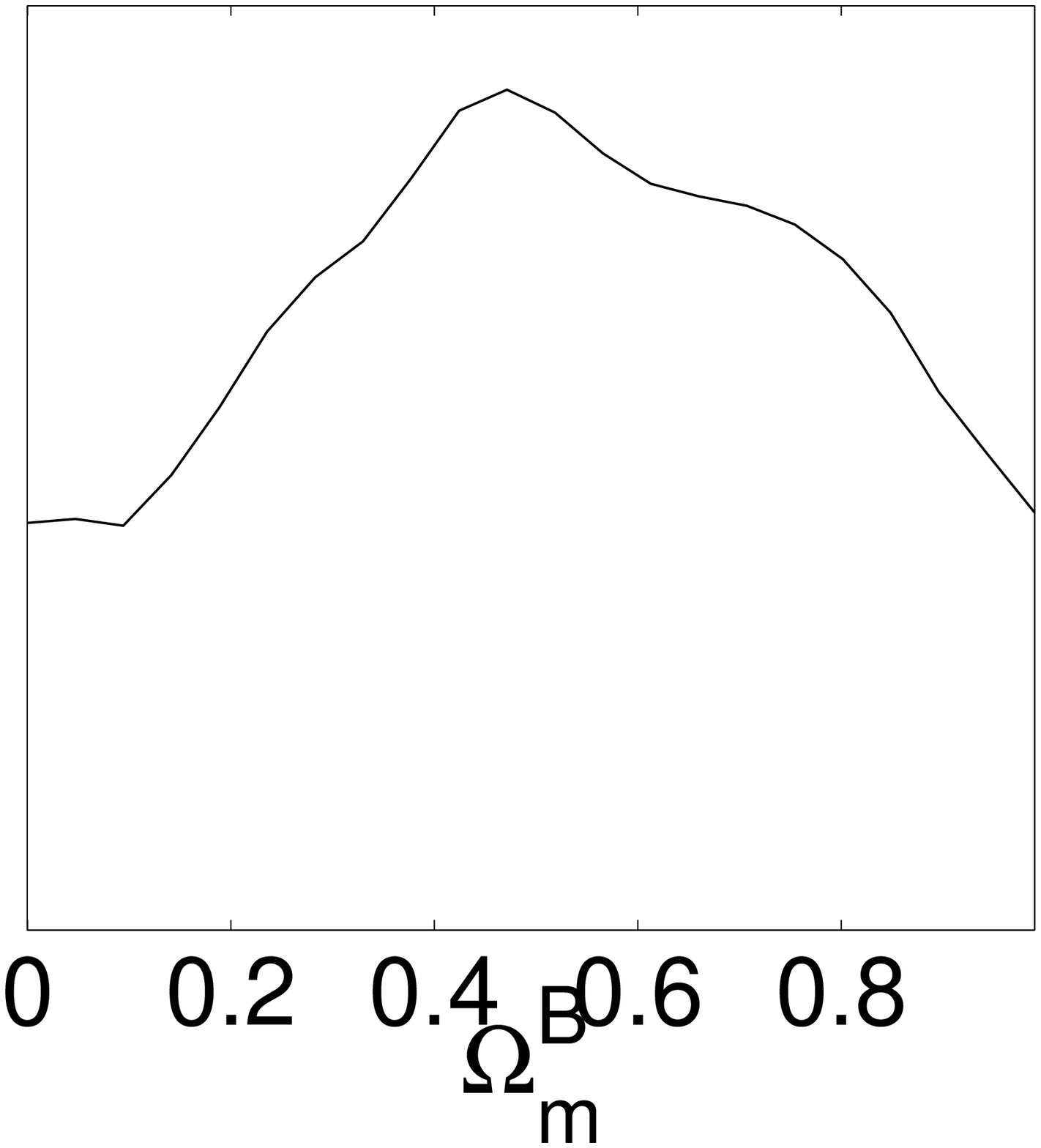}}\\
	\hfill
	  \subfigure{
           \includegraphics[width=.3\columnwidth]{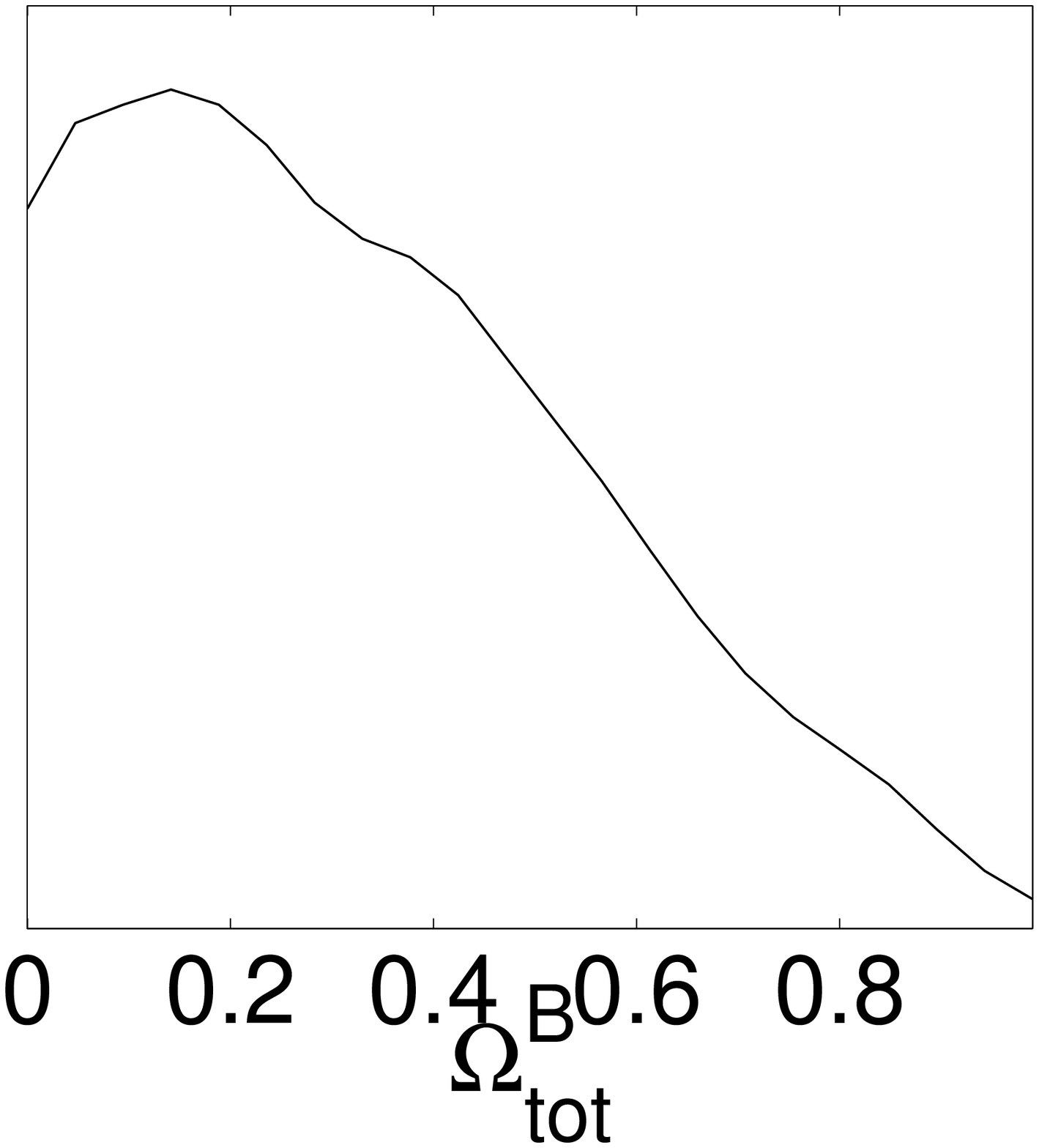}} 
\caption{Extracted Bianchi ${\rm VII_h}$ parameters including $\Omega_{\Lambda}^{\rm B}$
 (left handed model C) from WMAP 3-year data.}
\label{full}
\end{figure}

\begin{figure}
    	\subfigure{
          \includegraphics[width=.3\columnwidth]{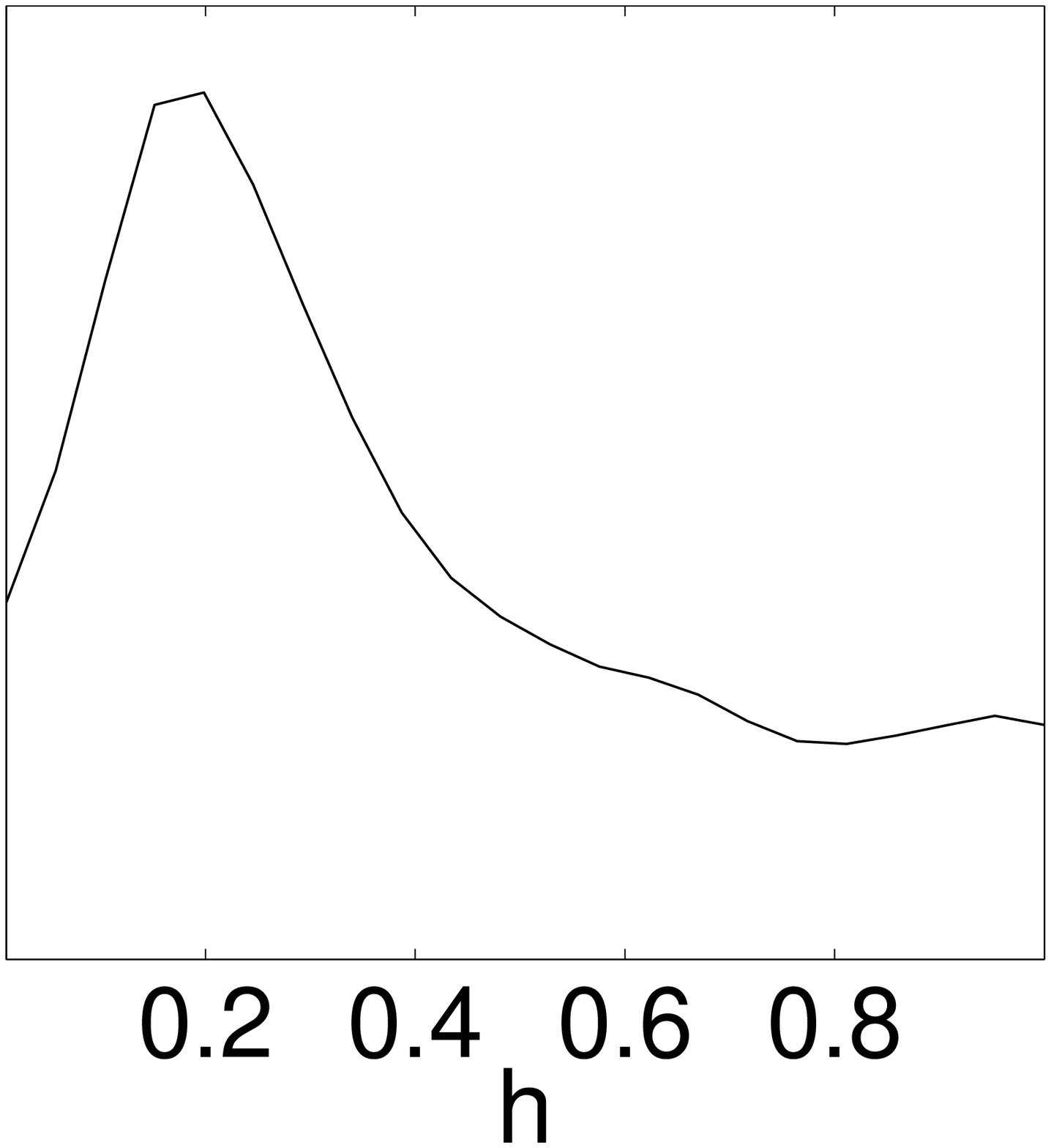}}
	\hfill
	\subfigure{
          \includegraphics[width=.31\columnwidth]{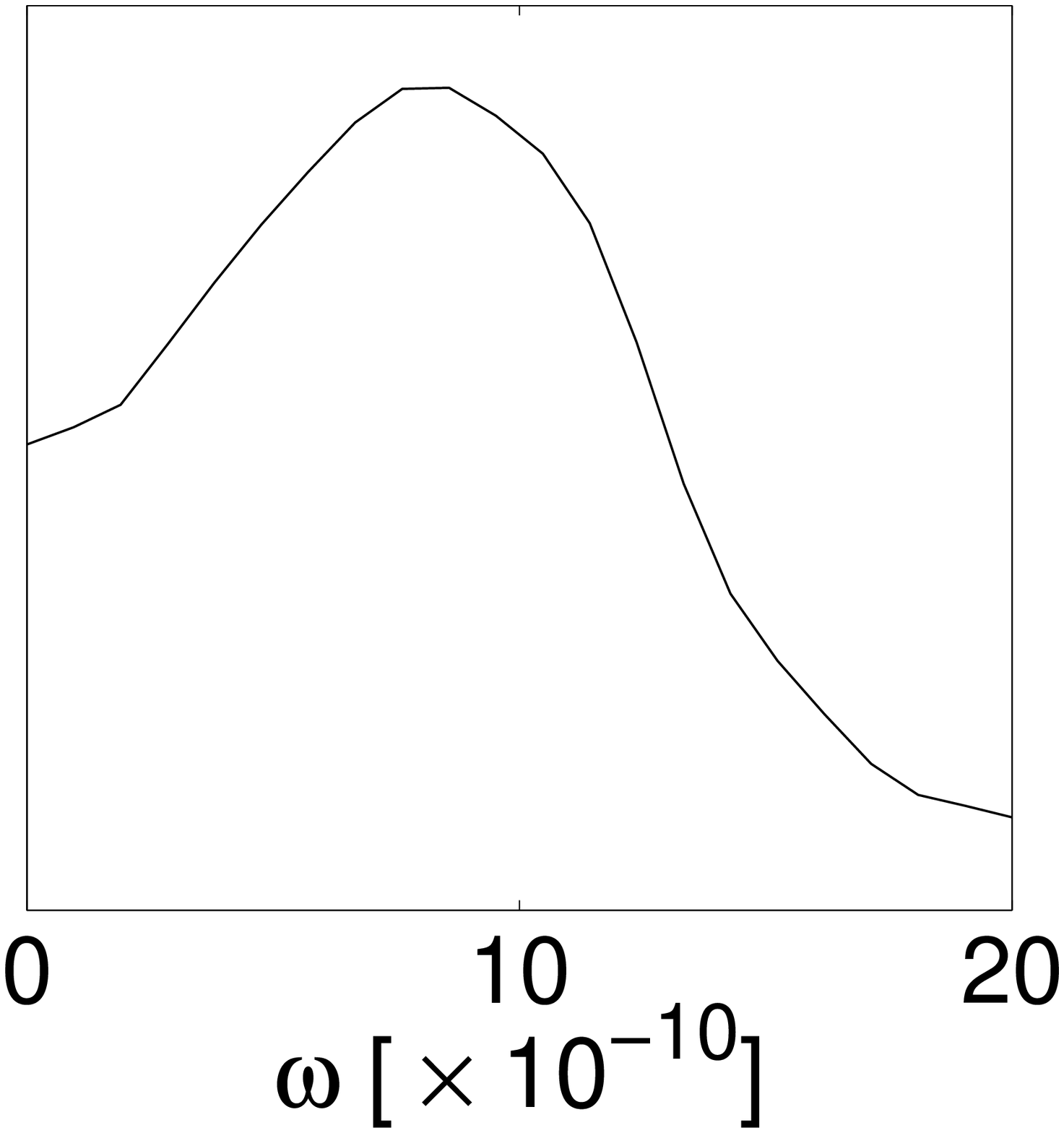}}
     	\hfill
	\subfigure{
           \includegraphics[width=.315\columnwidth]{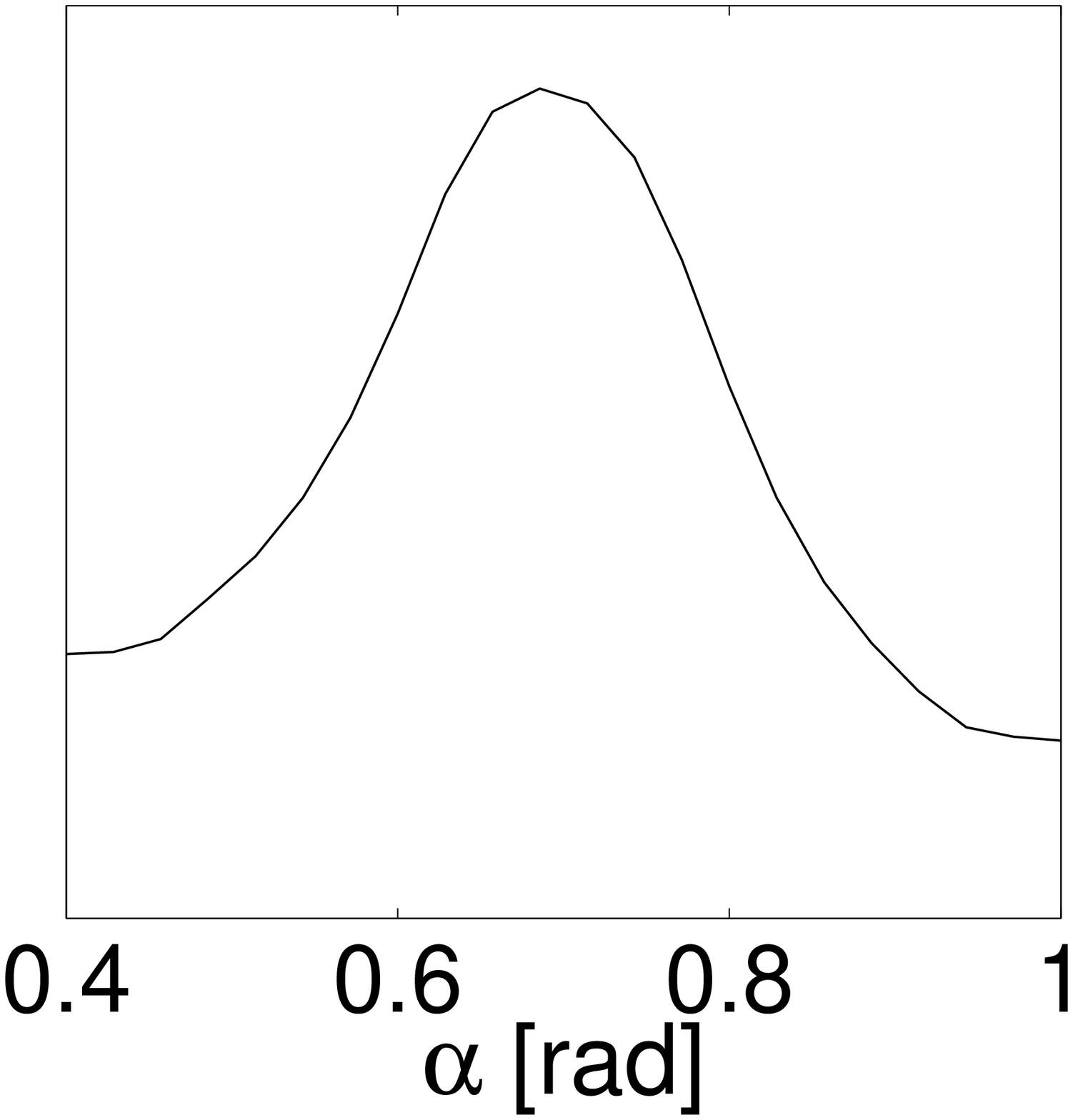}}\\
	\hfill
	\subfigure{
           \includegraphics[width=.31\columnwidth]{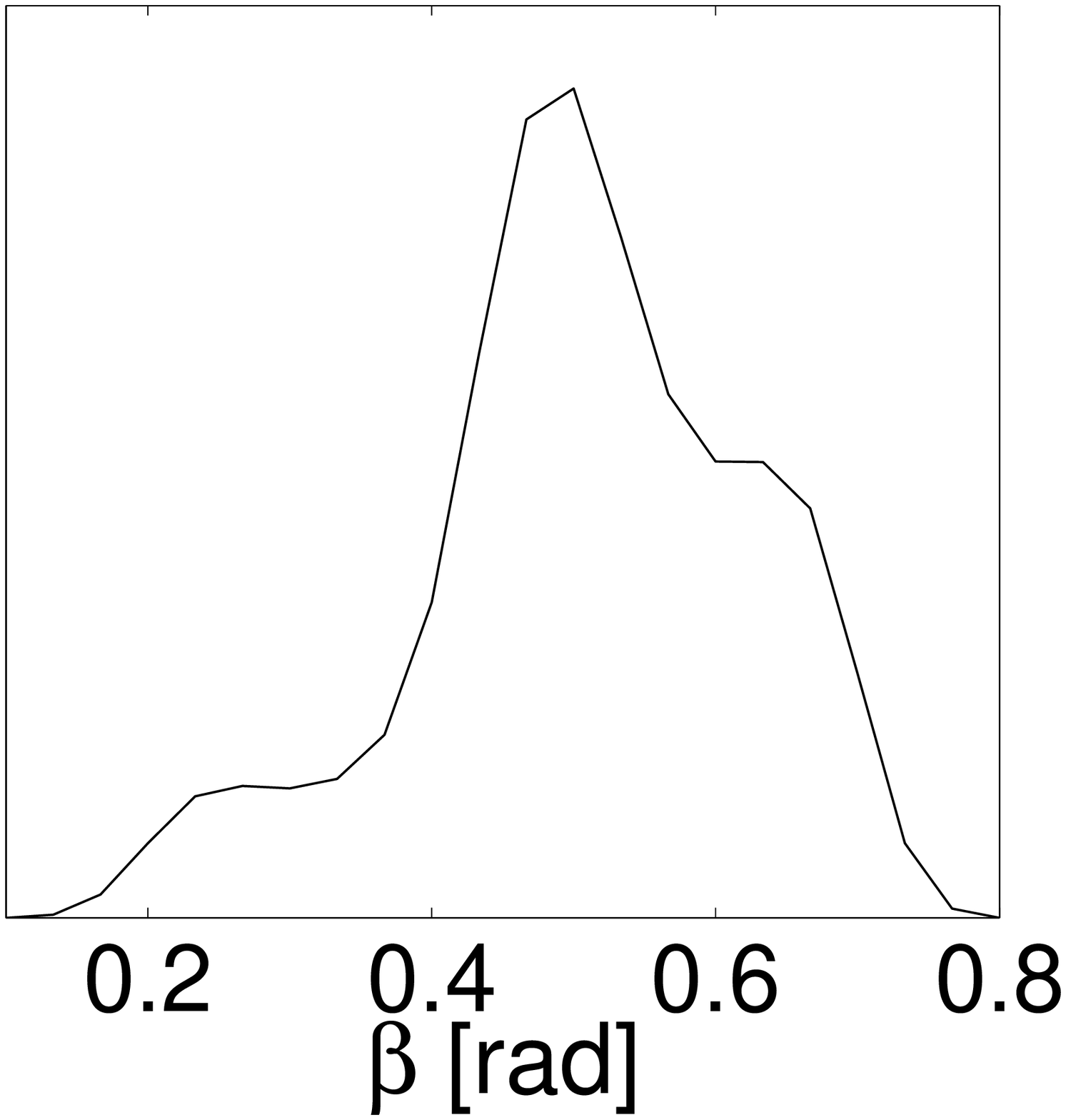}}
	\hfill
	\subfigure{
           \includegraphics[width=.3\columnwidth]{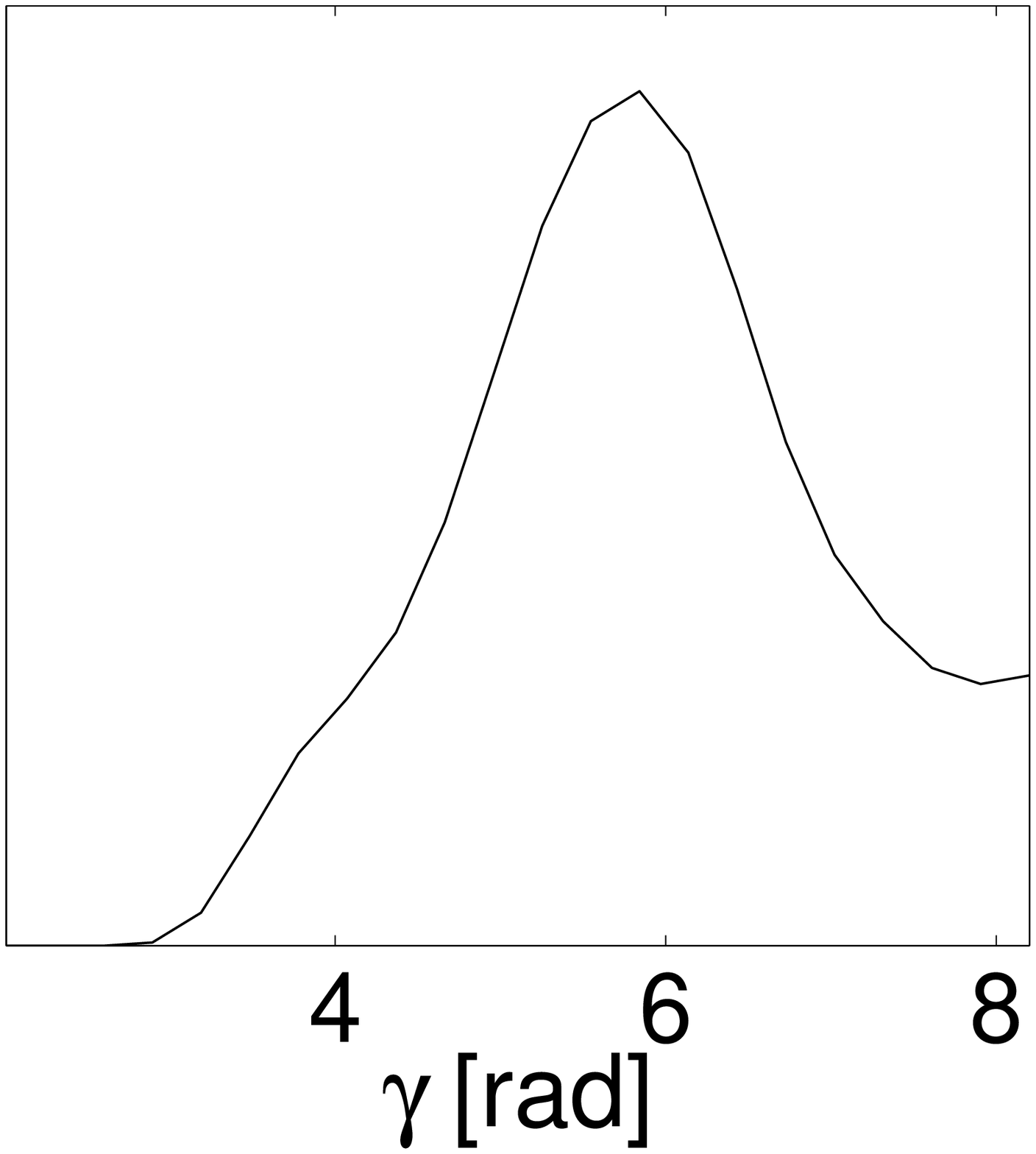}}
	\hfill
	\subfigure{
           \includegraphics[width=.3\columnwidth]{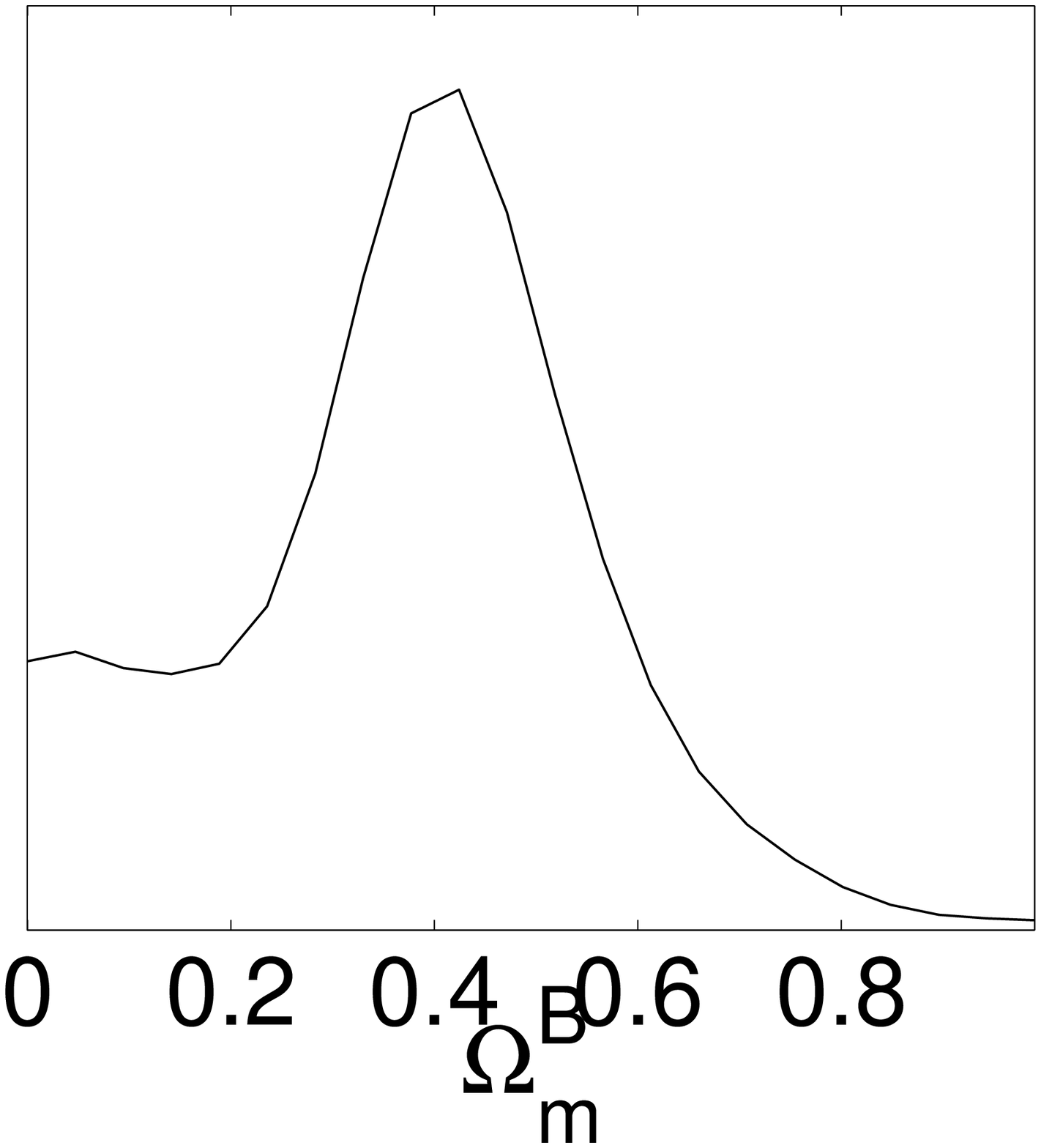}}
\caption{Extracted Bianchi ${\rm VII_h}$ parameters with a prior of $\Omega_{\Lambda}^{\rm B} = 0$
 (left handed model F) from WMAP 3-year data.} 
\label{reduced}
\end{figure}

Before continuing it is worth discussing in some detail the effect of each parameter on the structure
and scale of the Bianchi templates. The morphology is most sensitive to $\Omega_m^{\rm B}$ and
$h$, which determine the `focus' and `tightness' (also to some degree the amplitude) of
the spiral, respectively. The introduction of dark energy in the form of a cosmological
constant energy density creates a correlation between the matter density as the
repulsive nature of the former de-focuses the spiral. This degeneracy is almost exact,
being broken only by the overall amplitude. Unfortunately, many other parameters can
influence the amplitude of the CMB anisotropies on this scale such as the amplitude of
primordial scalar (and tensor) density perturbations, the optical depth to reionisation,
the spectral `tilt' of the primordial perturbations and of course the Bianchi vorticity
$\omega$. This profusion of possible (and mostly degenerate) explanations for the CMB
amplitude on large scales make constraints difficult to
obtain. Given these difficulties we feel it is justified to fix the cosmological
parameters at their WMAP 3-year best fit values \citep{Spergel} varying only the scalar
amplitude $A_s$. Moreover, careful choice needs to be made over the priors placed on the
Euler angles particularly as large regions of the ILC maps contain residual Galactic
noise and other emission. In this light we believe a reasonable restriction
on the position ($\alpha$ and $\beta$) of the centre of the pattern is well motivated,
while leaving the orientation angle ($\gamma$) free to rotate the spiral. Parameter constraints
with and without $\Omega_{\Lambda}^{\rm B}$ (models C and F) are shown in Figs \ref{full} and \ref{reduced} for a left-handed Bianchi
component. The angular ranges used, $\alpha = [0.4,1.0]$ and $\beta = [0.1, 0.7]$ do
implicitly place a prior on the centre of the feature being at least partially
associated with the `cold spot' in the southern hemisphere located at roughly $\alpha =
0.7$, $\beta = 0.4$. Right-handed models examined were entirely unconstrained by the
data (see Fig. 9) with a typical best fit
$\ln$ likelihood at least 2 units below left-handed models, 
substantiating the claim in \citet{Jaffec} that the only viable Bianchi fit is left-handed.  
The degenerate effect of $\Omega_{\Lambda}^{\rm B}$ is clearly evident in
the laxer constraints, and since we are now only assessing the feature as a non-physical
effect there is no benefit in retaining the parameter --it provides no additional effect
on the template structure. The above constraints provide robust confirmation of a template 
very similar to that found by \citep{Jaffea}.

\begin{figure}
    	\subfigure{
          \includegraphics[width=.3\columnwidth]{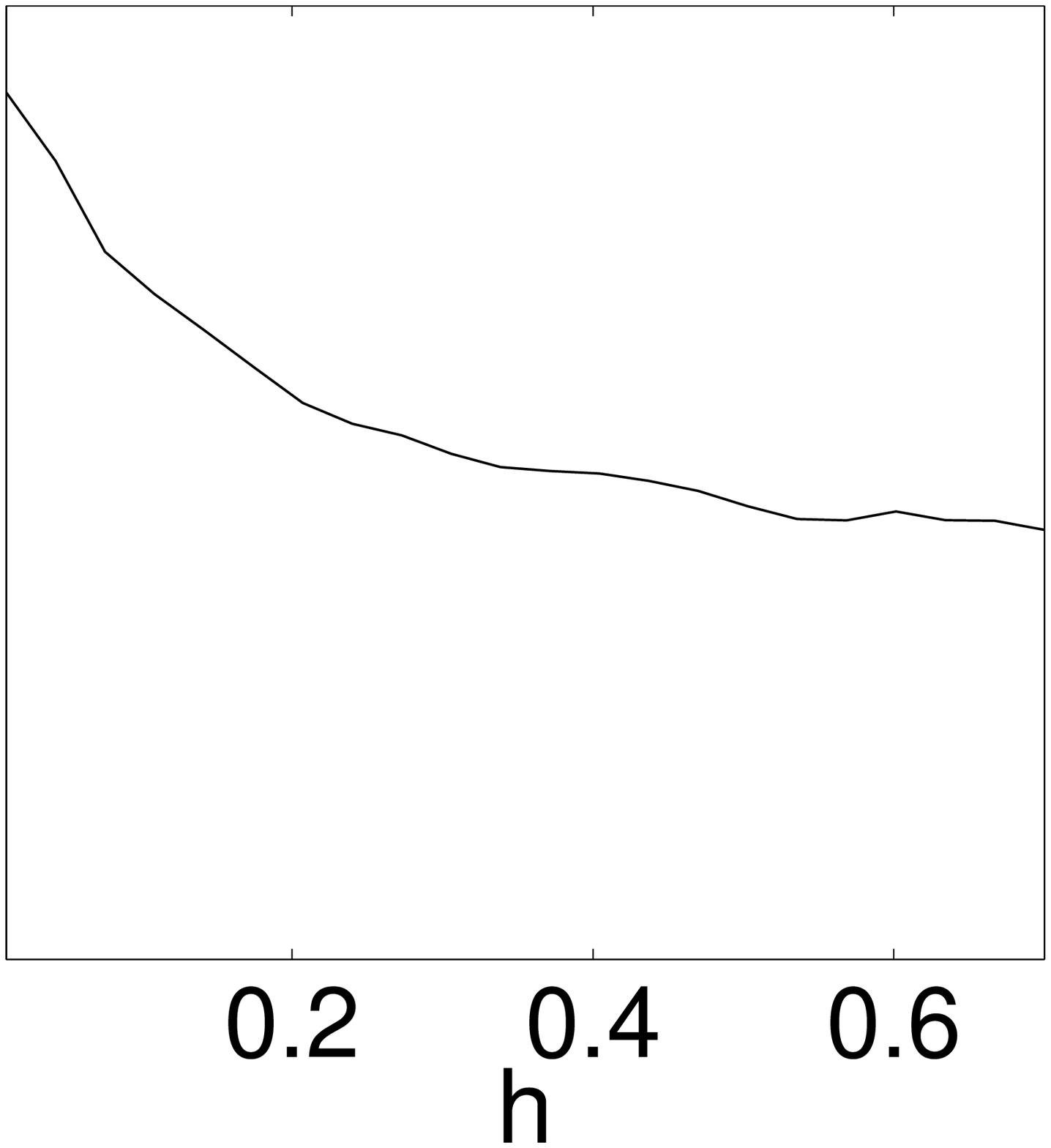}}
	\hfill
	\subfigure{
          \includegraphics[width=.31\columnwidth]{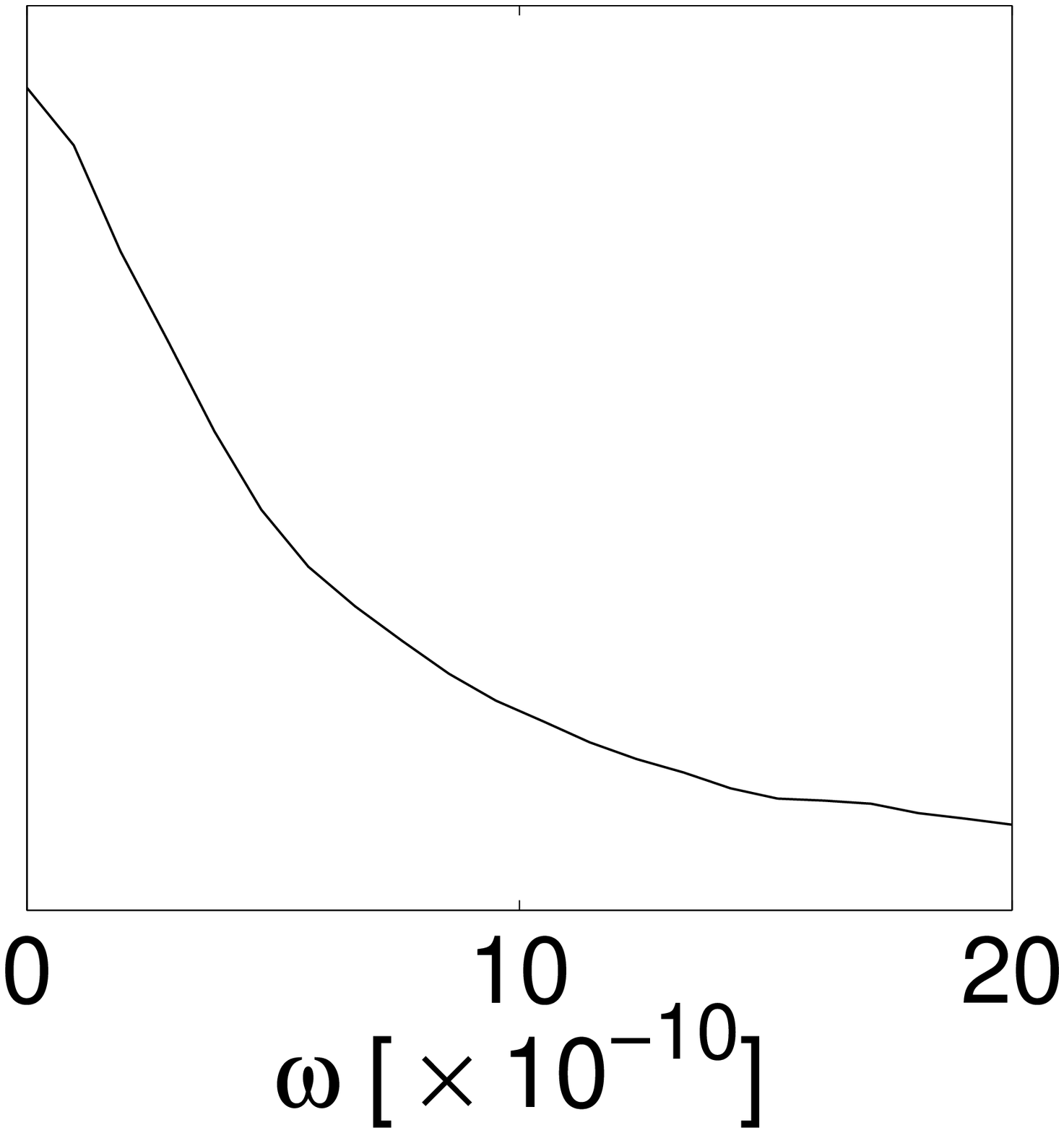}}
     	\hfill
	\subfigure{
           \includegraphics[width=.3\columnwidth]{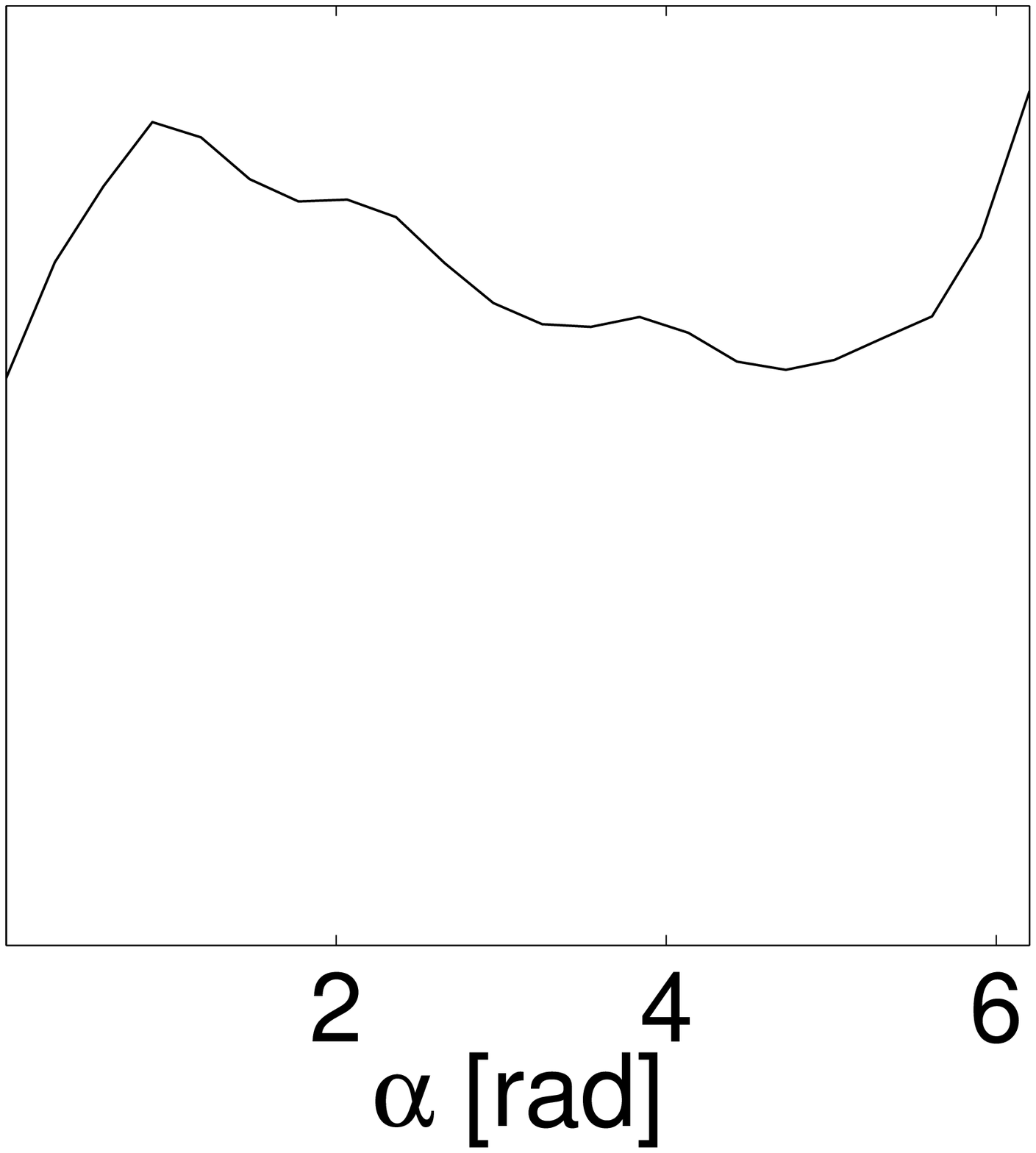}}\\
	\hfill
	\subfigure{
           \includegraphics[width=.3\columnwidth]{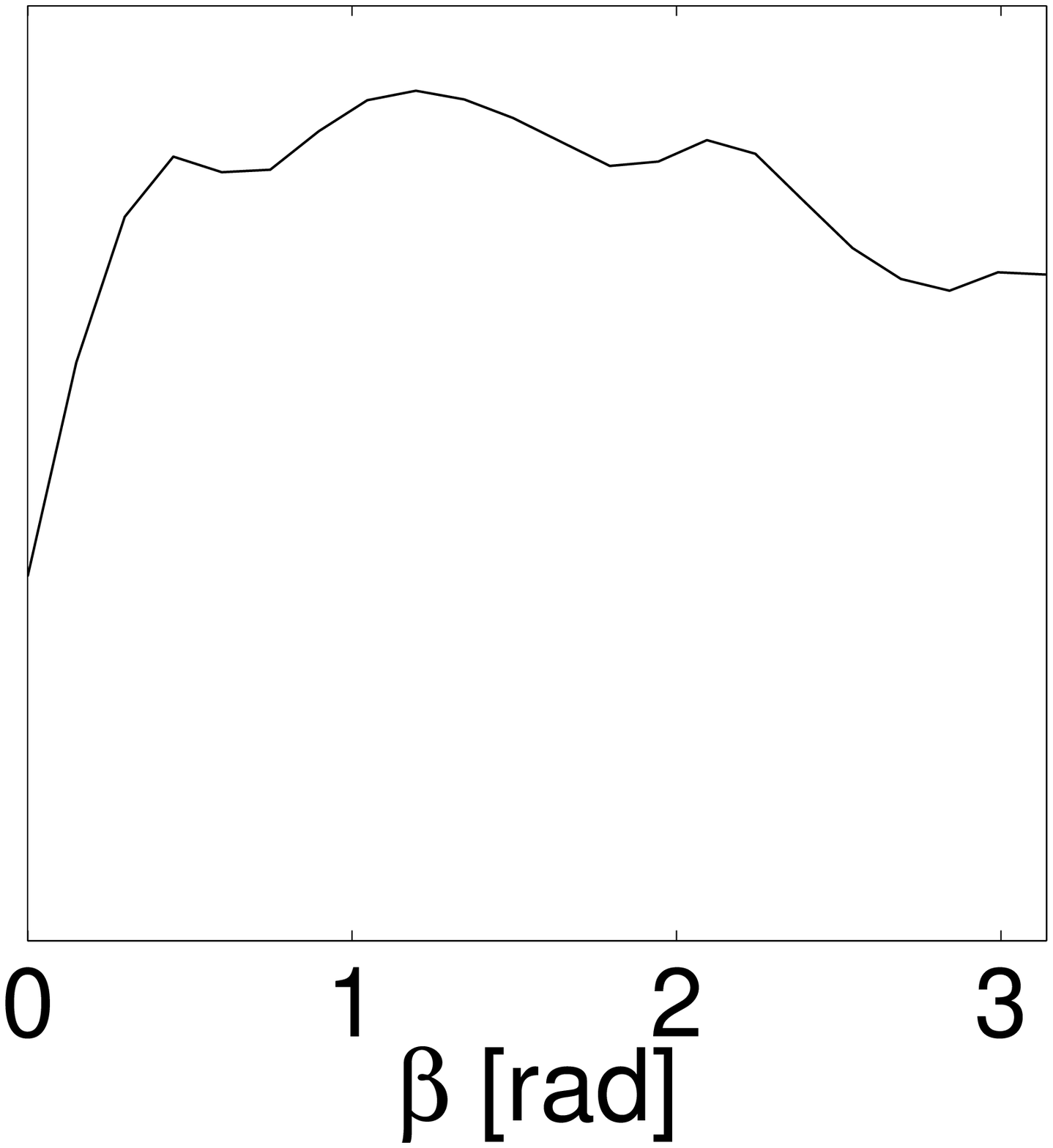}}
	\hfill
	\subfigure{
           \includegraphics[width=.3\columnwidth]{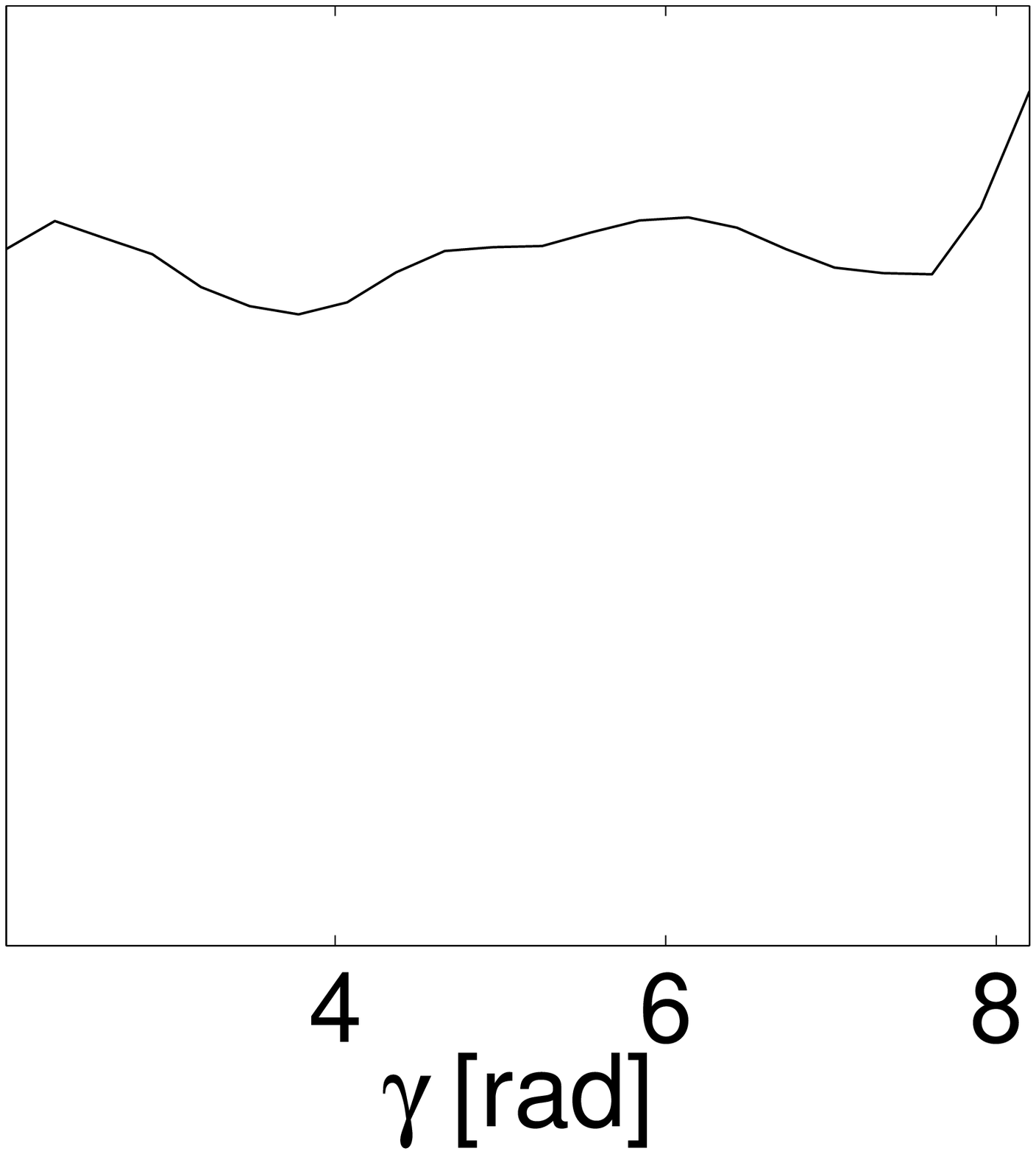}}
	\hfill
	\subfigure{
           \includegraphics[width=.3\columnwidth]{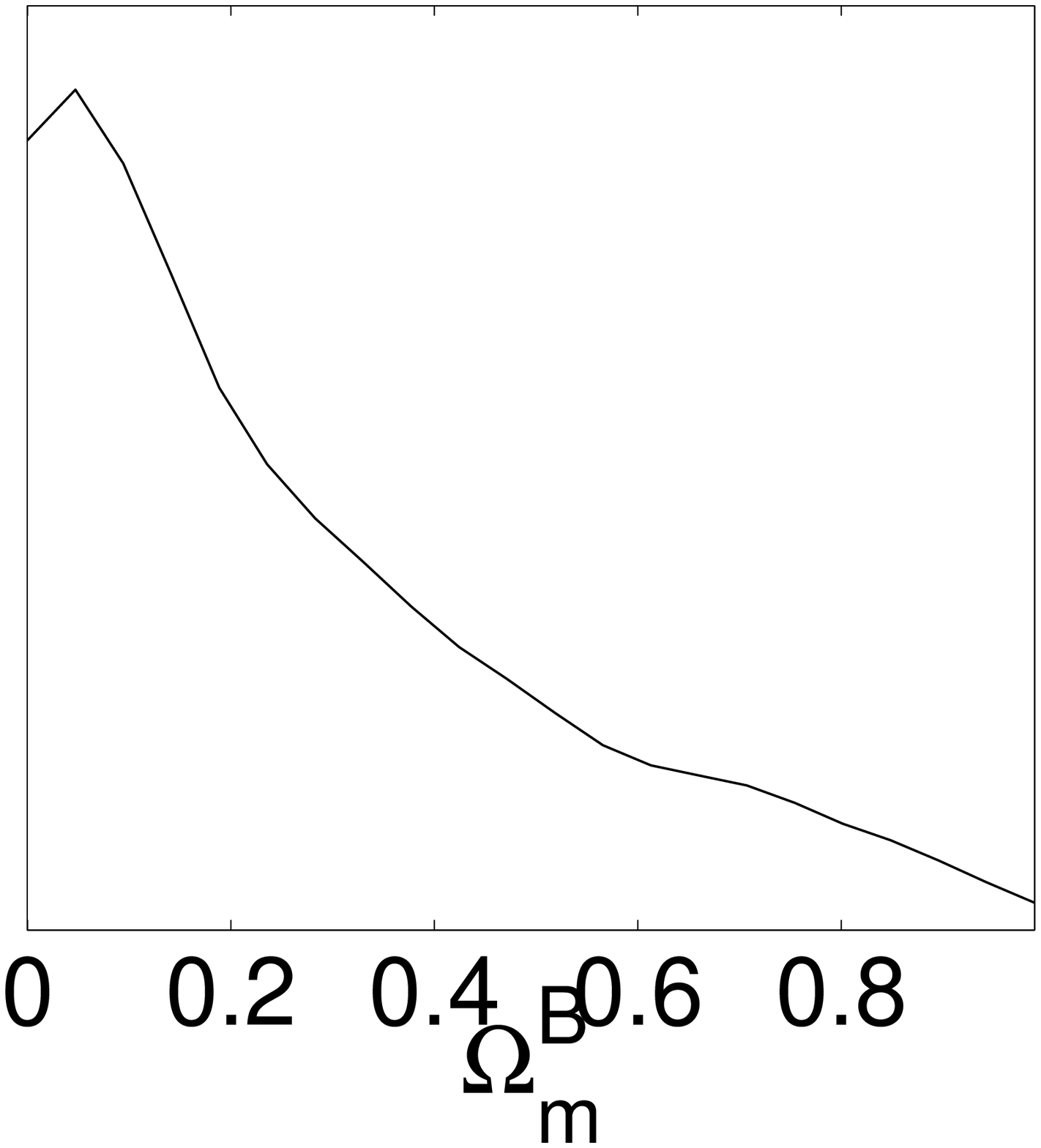}}
\caption{Extracted Bianchi ${\rm VII_h}$ parameters with a prior of $\Omega_{\Lambda}^{\rm B} = 0$ (right handed
model B) from WMAP 3-year data.} 
\label{Rhanded}
\end{figure}

\begin{figure}
    	\subfigure{
          \includegraphics[width=.3\columnwidth]{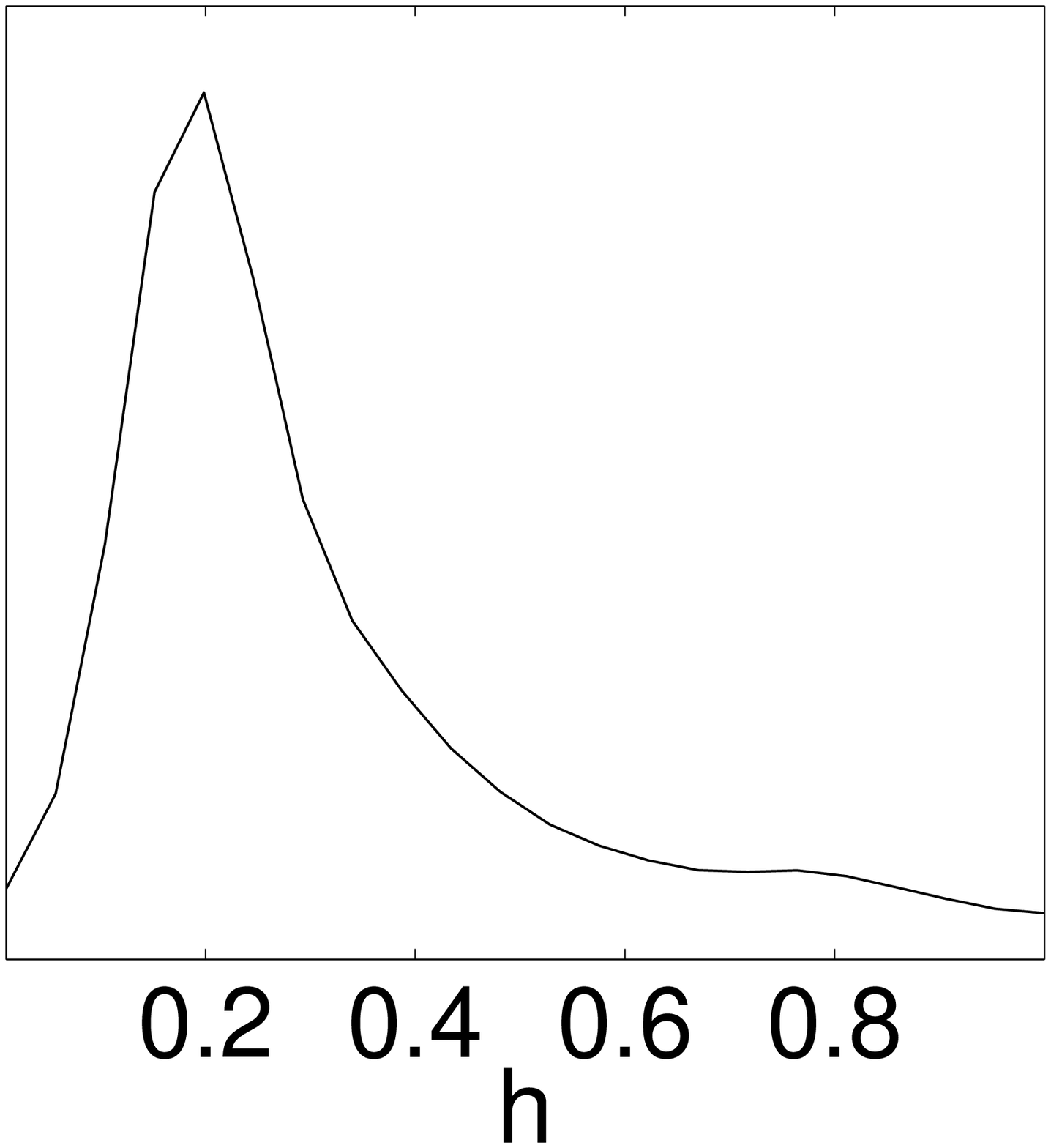}}
	\hspace{0.2cm}
	\subfigure{
          \includegraphics[width=.32\columnwidth]{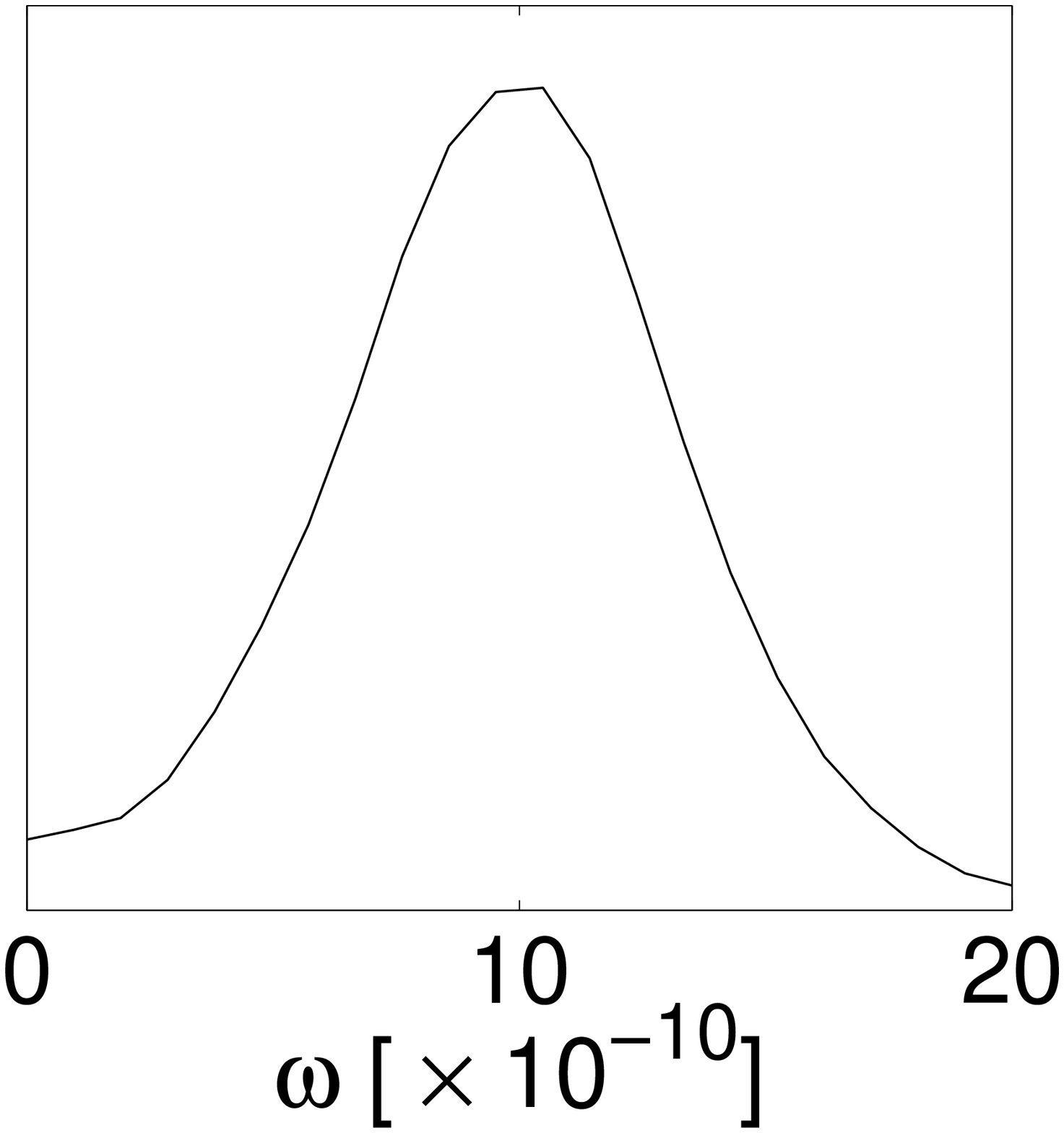}}\\
     	\subfigure{
           \includegraphics[width=.3\columnwidth]{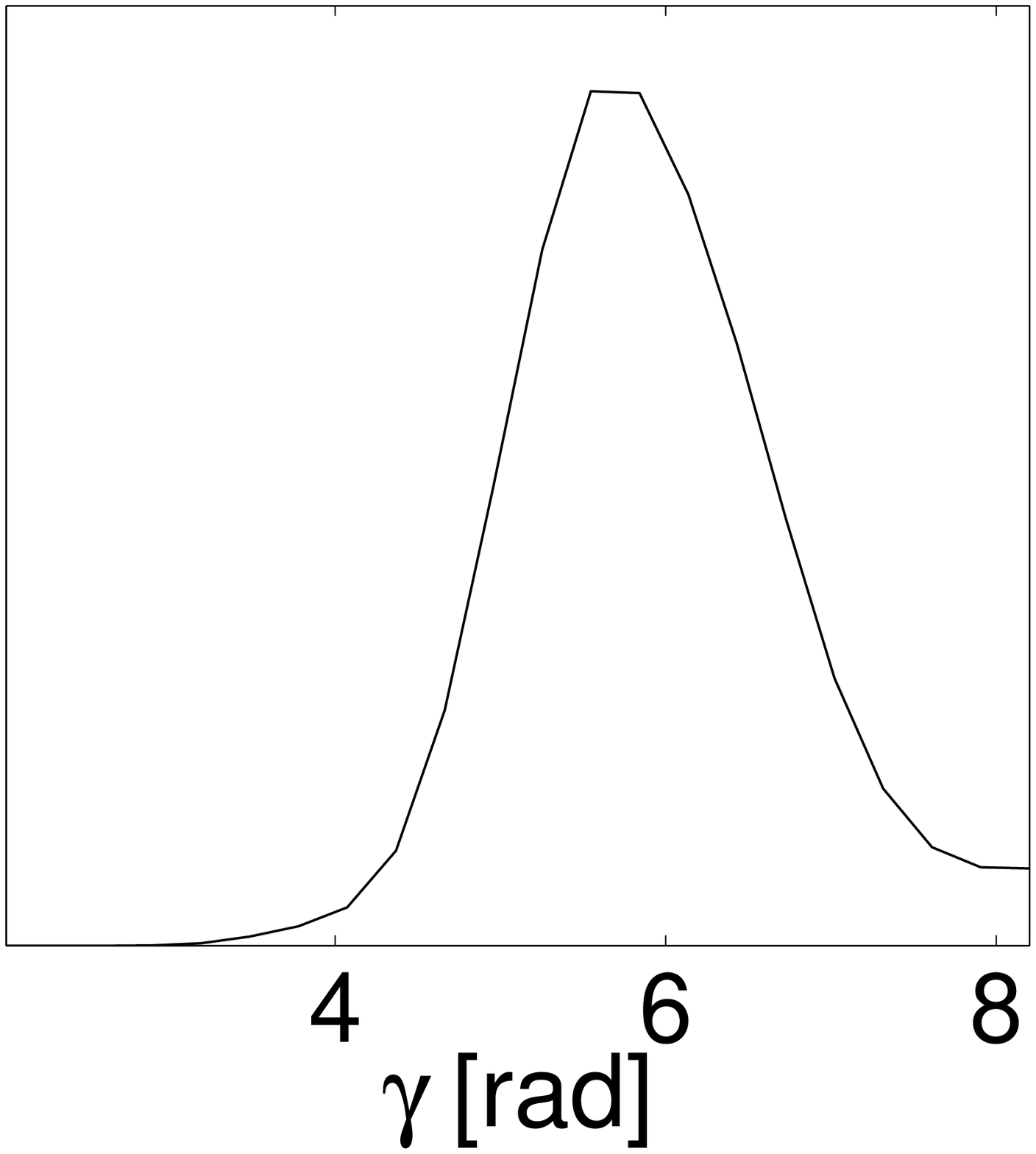}}
	\hspace{0.2cm}
	\subfigure{
           \includegraphics[width=.3\columnwidth]{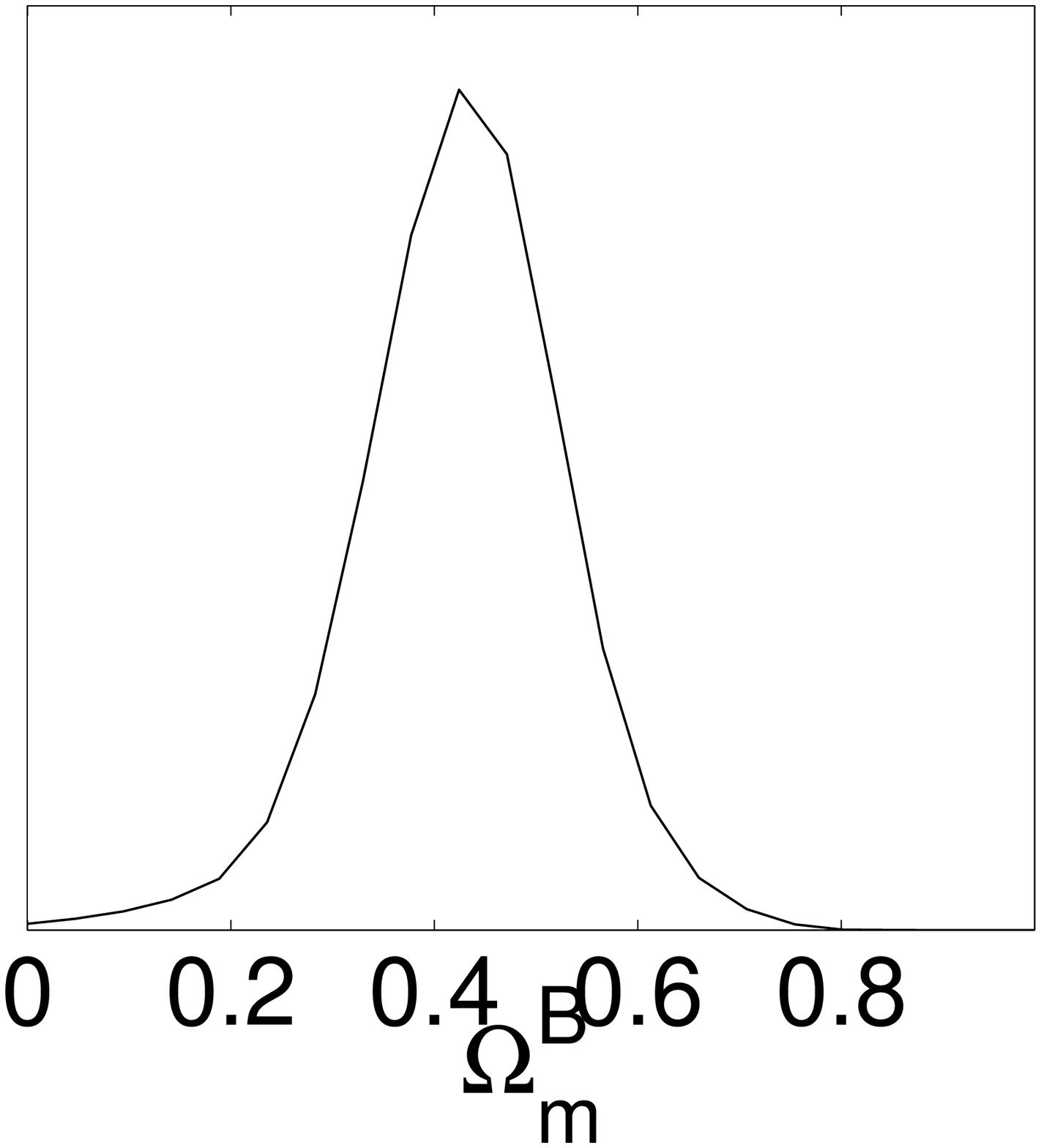}}
\caption{Extracted Bianchi ${\rm VII_h}$ parameters with a prior of $\Omega_{\Lambda}^{\rm B} = 0$ and centered on the `cold spot' (left handed
model G) from WMAP 3-year data. The only model considered which yields a positive evidence.} 
\label{restricted}
\end{figure}

\subsection{Model Selection}
\label{Model Selection}
The conclusion that the Bianchi ${\rm VII_h}$ models may only be treated as templates
does not preclude a consideration of the necessity of such a template to explain the
data. Indeed, removal of areas of non-Gaussianity and correction of large-scale power
would explain serious anomalies in the current data, but at the cost of removal of
universal isotropy. The Bayesian evidence provides an ideal way to assess quantitatively
whether such action is necessary. As with the previous section we shall assess a
template both with and without a dark energy density and with a range of priors on the
two positional Euler angles via models B -- G. All model comparisons are depicted via $\ln$ evidence differences; $\ln_{\rm A} - \ln_{\rm B-G}$ so that  
a positive value illustrates a preference for the inclusion of the Bianchi component and vice versa.

\begin{center}
\begin{table*}
\hspace{5cm}
\caption{$\ln$ evidence differences using one and three year all sky maps ILC maps for L and R handed models including a dark energy density.}
\begin{tabular}{|c||c||c||c||c||c||c|}
    \hline
 \textbf{Data} $\backslash$ \textbf{Model} 	& B (L)		&B(R)		&C(L)		 & C (R)	 & D (L)	 & D (R)	\\
    \hline
 3-year ILC 					& -0.9$\pm$ 0.2 & -1.4 $\pm$ 0.2& -0.6 $\pm$ 0.2 & -1.2 $\pm$ 0.2 & +1.2 $\pm$ 0.2& -1.0 $\pm$ 0.2\\
 1-year ILC					& -0.9$\pm$ 0.2 & -1.5 $\pm$ 0.2& -0.5 $\pm$ 0.2 & -1.1 $\pm$ 0.2 & +1.2 $\pm$ 0.2& -1.0 $\pm$ 0.2\\
     \hline
 \textbf{Data} $\backslash$ \textbf{Model} 	& E (L)		&E(R)		&F(L)		 & F (R)	 & G (L)	 & G (R)	\\
 	\hline    
 3-year ILC 					& -0.9$\pm$ 0.2 & -1.3 $\pm$ 0.2& -0.6 $\pm$ 0.2 & -1.1 $\pm$ 0.2 & +1.1 $\pm$ 0.2& -1.0 $\pm$ 0.2\\
 1-year ILC					& -0.8$\pm$ 0.2 & -1.5 $\pm$ 0.2& -0.7 $\pm$ 0.2 & -1.1 $\pm$ 0.2 & +0.9 $\pm$ 0.2& -1.0 $\pm$ 0.2\\
     \hline 
\end{tabular}
\label{evidence}
\end{table*} 
\end{center}

In agreement with the parameter constraints, little evidence exists for any of the right-handed
models, typically at least 1 $\ln$ evidence unit lower than for the left-handed models. We found
the results were relatively insensitive to the choice of priors on the energy densities, $h$ and
$\omega$,  but heavily dependant on the Euler angles. The only preferred model (a marginally
significant $\ln$ evidence difference $>$ 1) was found by restricting the pattern to lie centred
at the  `cold-spot' (models D and G), parameter constraints are shown in Fig. \ref{restricted}.
It must be noted that the Bayesian evidence is a prior dependent quantity, so that in
general a smaller prior will tend to lift the evidence and vice versa, thus an increased 
detection using these restricted priors is understandable.   
However, by still allowing the pattern to rotate freely, via $\gamma$ we still gain all
the degrees of freedom necessary to allow the data to decide the best morphology. Indeed all we
are saying with this prior is that we believe the `cold-spot' to be driving any detection. The
effect of increasing the parameter space with the addition of $\Omega_{\Lambda}^{\rm B}$ is
clealy offset by the large volumes of high likelihood that exist along the  degeneracy, leaving
the evidences almost unchanged. The conclusions are essentially independent on going from 1 to 3
year data, presumably as both maps are effectively cosmic variance limited up to at least $l<10$
where the majority of Bianchi structure lies.  

\section{Conclusions} 
We have investigated the possibility of signatures of universal shear and vorticity in 1
and 3-year WMAP data. We implemented Bianchi ${\rm VII_h}$ simulations in a fully
Bayesian MCMC analysis to extract the best fitting parameters \footnote{Our
best fitting Bianchi sky maps for 1 and 3 year data are shown in Appendix A and can be
downloaded in {\sc fits} format from http://www.mrao.cam.ac.uk/jdm57}. of the model and to
determine whether its inclusion is warranted by the data. We only make a marginally significant
detection for a left-handed Bianchi template with parameters; $\Omega_{tot} = 0.43\pm 0.04$, $h = 0.32^{+0.02}_{-0.13}$,
$\omega = 9.7^{+1.5}_{-1.6}\times 10^{-10}$ with orientation $\gamma = {337^{\circ}}^{+17}_{-21}$) if we restrict its position on the sky
to lie centred on the `cold-spot' at ($\alpha = 42^{\circ}$,$\beta = 32^{\circ}$) --all other models are essentially 
disfavoured by the data in both 1 and 3 year maps. Since the open universal geometry implied by these templates effectively
rules out the Bianchi ${\rm VII_h}$ models as the origin of this structure we can only
treat these models as template fittings, morphologically similar to some as yet unknown
physical cause. Nevertheless our analysis demonstrated that there are interesting effects on large scales yet to be explained in the WMAP data.

\section*{Acknowledgements}
This work was carried out largely on the COSMOS UK National
Cosmology Supercomputer at DAMTP, Cambridge and we would like to thank S. Rankin and V. Treviso for their computational assistance. The authors would like to 
thank David MacKay for useful comments. MB
was supported by a Benefactors Scholarship at St. John's College, Cambridge and an Isaac Newton Studentship. JDM was supported by a
Commonwealth (Cambridge) Scholarship.

\appendix
\section{Bianchi ${\rm VII_h}$ templates}
We illustrate here our best fitting Bianchi $\rm VII_h$ templates found from 3 year WMAP data. 
\begin{figure}
\begin{center}
\includegraphics[width=\linewidth]{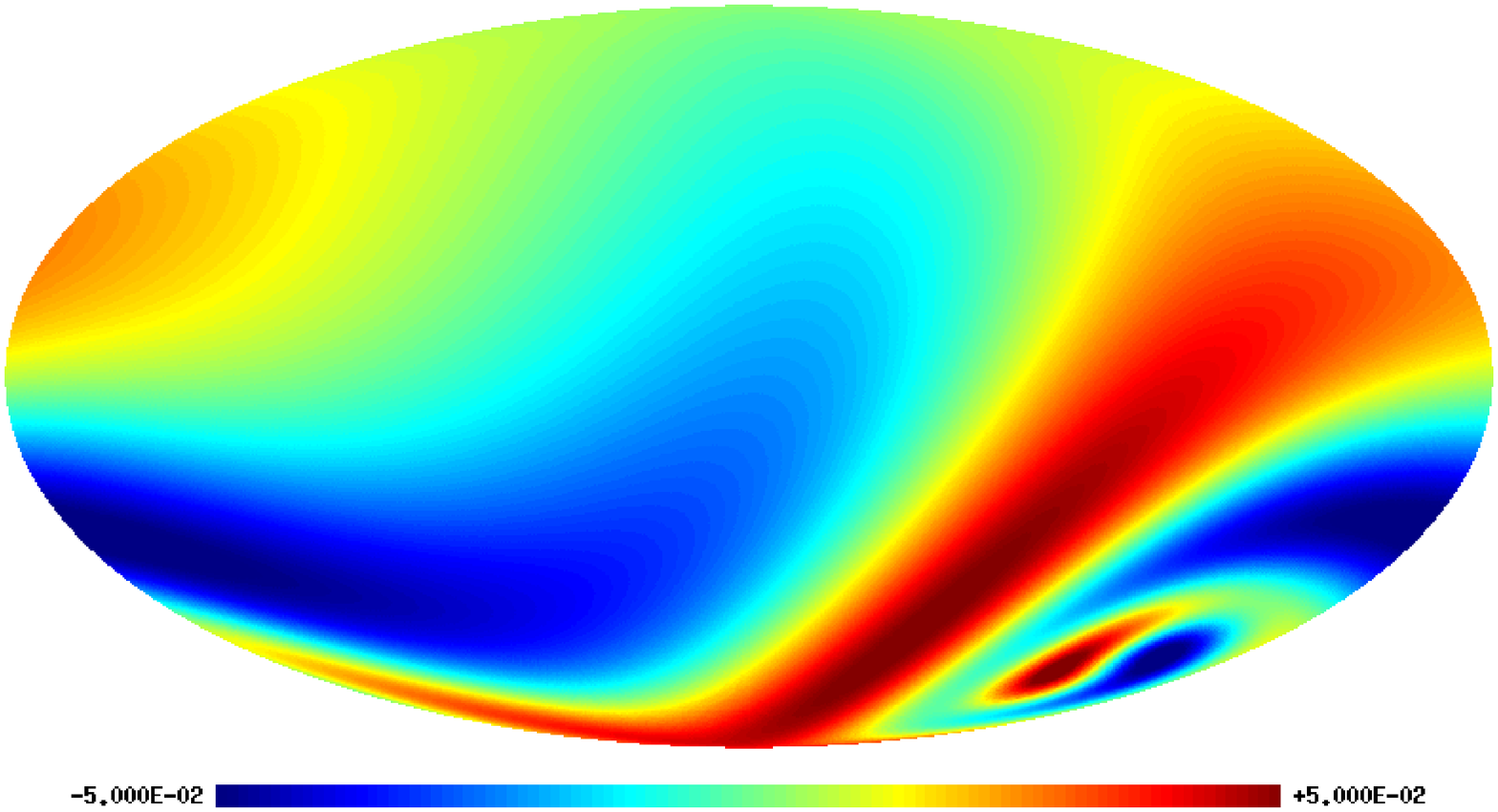}
\caption{Best fit bianchi template from WMAP 3 year data with parameters 
$\Omega_{tot} = 0.43\pm 0.04$, $h = 0.32^{+0.02}_{-0.13}$,
$\omega = 9.7^{+1.6}_{-1.5}\times 10^{-10}$ with orientation $\gamma =
{337^{\circ}}^{+17}_{-23}$) and position ($42^{\circ}$, $32^{\circ}$) in Euler angles.}
\label{wmap3 best fit}
\end{center}
\end{figure}

\label{lastpage}

\end{document}